\documentclass[ashowpacs,preprintnumbers,superscriptaddress,aps,prd,nofootinbib,floatfix,11pt]{revtex4-2}

\usepackage[T1]{fontenc}
\usepackage[utf8]{inputenc}
\usepackage{amsmath,amssymb}
\usepackage{booktabs}
\usepackage{multirow}
\usepackage{graphicx}
\usepackage{dcolumn}
\usepackage{float}
\usepackage{placeins} 
\usepackage{lmodern}
\usepackage{microtype}
\usepackage{xurl}
\usepackage{url}
\usepackage{hyperref}

\usepackage{caption}

\usepackage{dcolumn}
\usepackage{bm}
\usepackage{physics,amsfonts}
\usepackage{latexsym}
\usepackage{color}
\usepackage{subcaption}
\usepackage{ulem}
\usepackage{mathrsfs}
\usepackage{yhmath}
\usepackage{cancel}
\usepackage{tensor}

\usepackage[table]{xcolor}
\usepackage{colortbl}
\usepackage{array}

\newcolumntype{P}[1]{>{\centering\arraybackslash}p{#1}}
\newcolumntype{Q}[1]{>{\raggedright\arraybackslash}p{#1}}

\definecolor{colHdr}{RGB}{60,100,170}
\definecolor{colSub}{RGB}{210,222,242}
\definecolor{colA}{RGB}{244,247,253}
\definecolor{colB}{RGB}{255,255,255}
\definecolor{colGrp}{RGB}{225,232,245}
\definecolor{colCvg}{RGB}{240,243,250}

\usepackage[usestackEOL]{stackengine}
\renewcommand{\arraystretch}{1.5}
\usepackage{titlesec}

\renewcommand{\thesection}{\arabic{section}}
\renewcommand{\thesubsection}{\arabic{subsection}}

\hypersetup{%
setpagesize=false,
 bookmarksnumbered=true,%
 bookmarksopen=true,%
 colorlinks=true,%
 linkcolor=blue,
 citecolor=magenta,
 urlcolor=blue
}

\DeclareUnicodeCharacter{039B}{\Lambda}

\titleformat{\section}
  {\normalfont\large\bfseries}{\thesection}{1em}{}
\titlespacing*{\section}
  {0pt}{1.ex plus .1ex minus .1ex}{0.ex plus .1ex minus .1ex}

\makeatletter
\renewcommand{\p@subsection}{\thesection.}
\renewcommand{\p@subsubsection}{\thesection.\thesubsection.}
\makeatother

\titleformat{\subsubsection}
  {\normalfont\itshape}{\thesection.\thesubsection.\thesubsubsection}{1em}{}
\titlespacing*{\subsubsection}
  {0pt}{1.ex plus .1ex minus .1ex}{0.ex plus .1ex minus .1ex}

\renewcommand{\thesubsection}{.\arabic{subsection}}

\begin{document}
\allowdisplaybreaks
\flushbottom
\title{\boldmath A Global Overview of Starobinsky Inflation beyond slow-roll with CMB-BAO data}

\author{Tarun}
\affiliation{Department of Physical Sciences, Indian Institute of Science Education and Research Berhampur, Berhampur 760003, Odisha, India}
\author{Tanmoy Modak}
\affiliation{Department of Physical Sciences, Indian Institute of Science Education and Research Berhampur, Berhampur 760003, Odisha, India}

\author{Bj\"orn Malte Sch\"afer}
\affiliation{Astronomisches Recheninstitut, Zentrum f{\"u}r Astronomie der Universit\"at Heidelberg, Germany}

\begin{abstract}
We perform a global analysis of the Starobinsky inflation using most recent CMB data from ACT and SPT together with DESI BAO measurements. By directly solving the equations of motion, we find relatively stronger constraints on the model parameters than those obtained under the slow-roll approximation from the baseline $\Lambda$CDM parameters $A_s$ and $n_s$. We translate these constraints into limits on the reheating temperature. We also discuss the dependence of the prior choices of the model parameters. Further, the correlation of the model parameters with the BAO parameter $r_d h$ is also studied. We further investigate the impact of the global analysis on the dimension-six $R^3$ modified Starobinsky model, deriving constraints on its additional parameters. Finally, we also provide parameter limits on the effective inflationary  Hubble slow-roll (HSR) and potential slow-roll (PSR) parametrization.
\end{abstract}

\maketitle
\vspace{2ex}

\hrule
\vspace{2ex}
 \tableofcontents
\vspace{2ex}
\numberwithin{equation}{section}
\hrule
\setlength{\parskip}{1\baselineskip}
\setlength{\parindent}{0pt}
\vspace{1ex}
\section{Introduction}\label{sec:intro}
The Cosmic Microwave Background (CMB) anisotropies measured by space-based WMAP~\cite{WMAP:2012nax} and Planck~\cite{Planck:2018jri} satellites provide exquisite probes for the paradigm of inflation~\cite{Starobinsky:1980te,Sato:1980yn,Guth:1980zm}. Features such as the acoustic peaks in the CMB and the nearly scale-invariant adiabatic primordial power spectrum are in excellent agreement with the predictions of inflation. The marginalized spectral index $n_s$ and tensor-to-scalar ratio $r$, measured within the baseline $\Lambda$CDM model from CMB data, provide a powerful test for inflationary models. The Planck 2018 legacy data~\cite{Planck:2018jri} excluded several inflationary models and found that Starobinsky, or $R^2$, inflation~\cite{Starobinsky:1980te,Starobinsky:1983zz,Vilenkin:1985md,Mijic:1986iv,Maeda:1987xf} was one of the best-fitting models. For Starobinsky-like single-field attractor models, $n_s \simeq 1-\frac{2}{N_*}$, where $N_*$ is the number of $e$-folds between the horizon exit of the pivot scale $k_{\rm ref}=0.05 \ {\rm Mpc}^{-1}$ and the end of inflation. In Starobinsky inflation and the single-field regime of the $R^2$-Higgs model, $N_* \simeq 50$--$60$, corresponding to $n_s \simeq \left[0.960,0.967\right]$.

The recent combined measurements from CMB ground-based observatories, the Atacama Cosmology Telescope (ACT)~\cite{AtacamaCosmologyTelescope:2025blo} and the South Pole Telescope (SPT)~\cite{SPT-3G:2025bzu}, along with Planck data and baryonic acoustic oscillation (BAO) data from the Dark Energy Spectroscopic Instrument (DESI), have shifted $n_s$ to a higher value than that observed in the Planck 2018 data. The ACT collaboration combined its DR6 data with Planck (including lensing) and BAO from DESI-DR2, finding $n_s = 0.9752 \pm 0.0030$ (denoted as P-ACT-LB2)~\cite{AtacamaCosmologyTelescope:2025blo}. Later, the SPT collaboration also released its result, $n_s = 0.9728 \pm 0.0027$, combining its SPT-3G D1 data with Planck, ACT, and DESI-DR2 (denoted as CMB-SPA+DESI)~\cite{SPT-3G:2025bzu}. This higher value of $n_s$ from both collaborations disfavors the baseline Starobinsky inflation at the $\sim 2\sigma$ level. However, intriguingly, the CMB datasets themselves are found to be consistent with each other, as follows: Planck: $0.9657 \pm 0.0040$, ACT DR6: $0.9682 \pm 0.0069$, SPT-3G D1: $0.951 \pm 0.011$, SPT+ACT: $0.9671 \pm 0.0058$, SPT+Planck: $0.9636 \pm 0.0035$, and the combination of Planck+SPT+ACT i.e. CMB-SPA: $0.9684 \pm 0.0030$, respectively~\cite{SPT-3G:2025bzu}.\footnote{Note that all these experiments consider both $T$ and $E$ polarization data, as well as lensing reconstruction, and the prior on $\tau_{\rm{reio}}$ is taken to be the same as in Planck PR4.} The higher value of $n_s$ places several single-field attractor models that were consistent with Planck data in tension with Starobinsky inflation~\cite{AtacamaCosmologyTelescope:2025blo,SPT-3G:2025bzu}. This tension can be softened for $N_* > 60$, although it may require a nontrivial post-inflationary reheating history within Starobinsky inflation (see also similar discussions~\cite{Aoki:2025wld,Gialamas:2025kef,Zharov:2025zjg,Haque:2025uis,Yogesh:2025wak,Gialamas:2025ofz,Addazi:2025qra,Pallis:2025nrv,Saini:2025jlc,Wolf:2025ecy,Wang:2025dbj,Piva:2025cqi,SidikRisdianto:2025qvk,Zharov:2025zjg,Ferreira:2025lrd,Ellis:2025ieh,Odintsov:2025jky,Oikonomou:2025htz,Park:2025upd,Odintsov:2025eiv,Pozdeeva:2025wsl,Modak:2025bjv,Modak:2025grj,Pineda:2025ubm}).

Most analyses of Starobinsky inflation have relied on constraints on the marginalized $A_s$ and $n_s$ inferred from the P-ACT-LB2 or CMB-SPA+DESI data combinations within the baseline $\Lambda$CDM model under the slow-roll approximation. The primary aim of this paper is to perform a global analysis of the Starobinsky model by directly solving the equations of motion, without assuming the slow-roll approximation, and to derive constraints on its parameters. We also investigate the parameter degeneracies, the prior dependence of the inferred constraints, and the implications for the reheating temperature.

In addition, we study the correlation between the Starobinsky model parameter and the BAO parameter $r_d h$, where $r_d$ is the comoving sound horizon at baryon decoupling and $h$ is the reduced Hubble parameter in light of the CMB-BAO tension~\cite{SPT-3G:2025bzu}. The CMB-BAO tension refers to the discrepancy between the values of the matter density parameter $\Omega_m$ and the combination $r_d h$ inferred from CMB and DESI BAO data, where $r_d$ is the comoving sound horizon at the baryon drag epoch and $h$ is the reduced Hubble parameter. The partial degeneracy between the scalar spectral index $n_s$ and the physical matter density $\omega_m\equiv\Omega_m h^2$ in CMB data induces a positive correlation between $n_s$ and the BAO parameter $r_d h$. Consequently, the tension between the CMB and DESI BAO inferences of $\Omega_m$ and $r_d h$, when the datasets are combined, drives the inferred value of $n_s$ towards higher values. We analyze how $r_d h$  is correlated with Starobinsky model parameters between different data combinations.

We also present constraints on the $R^3$-modified Starobinsky inflation model. As in the baseline Starobinsky model, we examine the prior dependence and parameter degeneracies of the results. The $R^3$ modification is among the simplest extensions that can accommodate a higher $n_s$~\cite{Kim:2025dyi,Addazi:2025qra,Modak:2025bjv}. For completeness, we also update the constraints on the Hubble slow-roll (HSR)~\cite{Lesgourgues:2007aa,Planck:2018jri} and potential slow-roll (PSR) parametrizations~\cite{Planck:2013jfk,Planck:2018jri} using these new datasets. These model-independent constraints can be translated and recast into a wide range of specific inflationary models.

The article is organized as follows. The details of different datasets and likelihoods are provided in Sec.~\ref{sec:datasets}: all CMB-related data are described in Sec.~\ref{subsec:cmbspa}, followed by BAO data from DESI in Sec.~\ref{subsec:bao}, and BICEP/Keck 2018 data in Sec.~\ref{subsec:bk18}. In Sec.~\ref{sec:Staro} and Sec.~\ref{sec:R3mod}, we present our results and inferences baseline and $R^3$ modified Starobinsky model. The constraints on model independent potential and Hubble slow-roll inflationary parameterizations are discussed in Sec~\ref{sec:hsr}. We conclude with a summary and outlook in Sec.~\ref{sec:disc}.

\section{Analysis framework, Datasets and Likelihoods}
\label{sec:datasets}
We implement the inflationary potentials, CMB angular power spectra and matter power in CLASS~\cite{Blas:2011rf,Lesgourgues:2011rg} which then interfaced with the Cobaya framework~\cite{Torrado:2020dgo} for Bayesian parameter inferences. The nonlinear corrections to the matter power spectrum for large-scale structure observables are included via the \textit{HaloFit} prescription while other cosmological parameters set to default values within the CLASS framework. The posterior sampling is performed using the Metropolis-Hastings MCMC algorithm implemented in Cobaya~\cite{Torrado:2020dgo}. The datasets individual likelihood components are described below. For the combination of the CMB data from Planck, ACT, and SPT, and BAO data from DESI, we closely follow the prescription given in Ref.~\cite{SPT-3G:2025bzu}.

\subsection{The CMB datasets}
\label{subsec:cmbspa}

\subsubsection{SPT-3G D1 likelihood.}
We use the SPT-3G first data release (SPT-3G D1) temperature and E-mode polarisation likelihood which is implemented via the \textsc{candl} interface~\cite{SPT-3G:2025bzu}. The dataset covers approximately $4\%$ of the sky ($\sim 1500~\mathrm{deg}^{2}$ of the southern sky). The measurements of the lensed EE and TE spectra are the most precise available at $\ell\approx 1800$--$4000$ and $\ell\approx 2200$--$4000$, respectively~\cite{SPT-3G:2025bzu}. As in Ref.~\cite{SPT-3G:2025bzu}, we use the ``lite'' variant of the \textsc{candl} likelihood, which is marginalized over the foreground and beam systematics at the band power level, while retains all cosmological sensitivity. In our numerical framework, we have removed the internal \textsc{candl} priors such that the prior information enters only through the Cobaya parameter specifications. The two nuisance parameters are included as free sampling variables. We apply a flat prior over the interval $[0.8, 1.2]$ for the polarisation calibration factor, $E_{\rm cal}$. For the temperature calibration factor, $T_{\rm cal}$, we impose a Gaussian prior of $\mathcal{N}(1.0, 0.0036)$ that is truncated to the same $[0.8, 1.2]$ range.

In addition, we include the SPT-3G CMB lensing and delensed EE power spectrum reconstruction through the \texttt{muse3glike} likelihood, based on the Marginal Unbiased Score Expansion (MUSE) method~\cite{SPT-3G:2025bzu}. MUSE is a Bayesian map-based inference approach, used because the standard quadratic estimator (Hu-Okamoto) reconstruction~\cite{Hu:2001kj,Okamoto:2003zw} becomes suboptimal at the depth of the SPT-3G D1 polarization maps, where the delensing is applied solely to the EE spectrum. This asymmetric treatment is primarily adopted because sharpening the acoustic peaks leads to the largest improvement in signal-to-noise for the EE spectrum, which directly improves the constraint on $A_s$. When combined with the $\phi\phi$ lensing potential auto-spectrum, this breaks the $A_s$-$\tau_{\rm{reio}}$ degeneracy (i.e. $A_s e^{-2\tau_{\rm{reio}}}$) degeneracy directly using lensing information. However, we have not applied delensing to the TT and TE spectra for all datasets, instead, we rely on standard forward modelling, where the CLASS Boltzmann solver computes the fully lensed theoretical spectra and compares them directly against the raw band powers.

\subsubsection{ACT DR6 primary CMB likelihood.}
We include the ACT Data Release 6 (ACT DR6) primary CMB likelihood~\cite{AtacamaCosmologyTelescope:2025blo,AtacamaCosmologyTelescope:2025nti}. The ACT DR6 maps cover $19{,}000~\mathrm{deg}^{2}$ of sky in three frequency bands centred at 98, 150, and 220~GHz, with white-noise levels in polarisation approximately three times lower than Planck~\cite{AtacamaCosmologyTelescope:2025blo}. Due to the overlapping sky coverage, we utilize the joint Planck and ACT (P-ACT) dataset in our analysis as described by~\cite{SPT-3G:2025bzu}. We incorporate this dataset through two complementary likelihood components in the Cobaya configuration. The first one is the \texttt{act\_dr6\_cmbonly.PlanckActCut}, which incorporates the joint P-ACT foreground-marginalised likelihood over the multipole ranges $\ell_{\rm TT}\in[0,1000]$, $\ell_{\rm TE}\in[0,600]$ and $\ell_{\rm EE}\in[0,600]$ using the plik lite likelihood to cover the high-$\ell$ Planck range (\texttt{plik\_lite\_v22} dataset). A single overall calibration parameter $A_{\rm ACT}$ is included as a free sampling variable; the Planck calibration is tied to it via $A_{\rm Panck}=A_{\rm ACT}$. The second component, \texttt{act\_dr6\_cmbonly}, covers the full high-$\ell$ ACT range $\ell_{\rm TT/TE/EE}\in[600,6500]$ accessing the publicly released \texttt{dr6\_data\_cmbonly.fits} SACC file~\cite{AtacamaCosmologyTelescope:2025blo}. This component introduces an additional polarisation efficiency parameter $P_{\rm ACT}$ with uniform prior $P_{\rm ACT}\in[0.9,1.1]$.

We further include the ACT DR6 CMB lensing reconstruction via the ACT DR6 LensLike likelihood, using the \texttt{actplanck\_baseline} variant at $\ell_{\rm max}=4000$. Using this we implement the combined lensing likelihood of ACT DR6 and Planck PR4 lensing~\cite{SPT-3G:2025bzu}. We have also applied the Hartlap correction to the covariance matrix.

\subsubsection{Planck likelihood.}
We have included Planck 2018 low-$\ell$ TT likelihood~\cite{Planck:2018vyg} in the range $\ell\in[2,29]$ to constrain the large-angle temperature anisotropies. The large-scale E-mode polarisation is incorporated through an informative Gaussian prior on $\tau_{\rm{reio}}$ derived from the Planck PR4 analysis as suggested in Ref.~\cite{SPT-3G:2025bzu}. This ensures our dataset combinations are robust against the anomalous lensing amplitude ($A_L \approx 1.18$) present in the full Planck dataset~\cite{Planck:2018vyg}. This anomaly is predominantly driven by excess acoustic peak smoothing between $\ell \sim 500$--$2500$ in the temperature spectrum. We mitigate this through two complementary mechanisms. First, the \texttt{PlanckActCut} pipeline strictly limits the Planck data to $\ell \leq 1000$ for TT and $\ell \leq 600$ for TE/EE, excising the multipole range where the anomalous smoothing signal is most statistically significant. Second, all temperature and polarisation information above these thresholds is supplied entirely by ACT DR6 and SPT-3G. Both high-resolution datasets are independently consistent with $A_L = 1$ in standard $\Lambda$CDM and dominate the constraining power at intermediate and small angular scales. Consequently, any residual anomalous $A_L$ pull from the retained Planck multipoles is statistically overwhelmed and does not bias our cosmological constraints.

\subsection{DESI DR2 and the CMB-BAO tension}
\label{subsec:bao}
We include the full DESI DR2 BAO likelihood~\cite{DESI:2025zgx} in our dataset and investigate its implications for the Starobinsky model without imposing the slow-roll approximation. As discussed above, the CMB-BAO tension arises primarily from the discrepancy between the matter density parameter $\Omega_m$ and the combination $r_d h$ inferred from CMB-only and DESI BAO datasets~\cite{SPT-3G:2025bzu}. Within the $\Lambda$CDM framework, the tension is reported to be approximately $3.1\sigma$ when using ACT data and increases to $3.7\sigma$ when ACT is combined with SPT data~\cite{SPT-3G:2025bzu}. Although the scalar spectral index $n_s$ is not directly constrained by BAO measurements, it can be indirectly affected through parameter degeneracies. Ref.~\cite{Ferreira:2025lrd} shows that $\Lambda$CDM fits to CMB data exhibit a strong positive correlation between $n_s$ and the BAO parameter $r_d h$, arising from the partial degeneracy between $n_s$ and the physical matter density $\omega_m (\equiv \Omega_m h^2)$. Thus, we also expect the inflationary model parameters in our analysis to show a similar feature.

\subsection{BICEP/Keck 2018}
\label{subsec:bk18}
The BICEP/Keck 2018 (BK18) B-mode polarisation likelihood~\cite{BICEP:2021xfz} which directly constrains the primordial tensor-to-scalar ratio $r$ and break the $n_{s}$--$r$--$\tau_{\rm{reio}}$ degeneracy from CMB T and E mode analyses. The BK18 likelihood is implemented via combining multi-frequency observations from the BICEP/Keck array at 95, 150, and 220~GHz with auxiliary Planck and WMAP polarisation maps at 23, 30, 33, 44, 143, 217, and 353~GHz via multi-frequency component separation~\cite{BICEP:2021xfz}. The B-mode auto- and cross-spectra are evaluated over $\ell\in[20,330]$, using the eleven map combinations
\begin{equation}
\begin{aligned}
  & \mathrm{BK18\_K95\_B}, \mathrm{BK18\_150\_B}, \mathrm{BK18\_220\_B}, \mathrm{BK18\_B95e\_B}, \\
  & \mathrm{W023e\_B}, \mathrm{P030e\_B}, \mathrm{W033e\_B}, \mathrm{P044e\_B}, \mathrm{P143e\_B}, \mathrm{P217e\_B}, \mathrm{P353e\_B}
\end{aligned}
\label{eq:BK18_maps}
\end{equation}
with multipole range \texttt{use\_min}$=1$ to \texttt{use\_max}$=9$ in the standard BK18 band-power bins.

Seven foreground nuisance parameters are sampled simultaneously with the cosmological parameters; their prior specifications (see Sec.~\ref{subsec:datacomb}) follow the official BK18 data release~\cite{BICEP:2021xfz}. Four geometric correlation parameters $\delta_{\rm dust}$, $\delta_{\rm sync}$, $\gamma_{\rm corr}$, $\gamma_{95},_{150},_{220}$ are set at the default values in Cobaya.

The inclusion of BK18 is essential for two reasons beyond the obvious tensor constraint. First, the current $95\%$ CL upper limit $r<0.036$~\cite{BICEP:2021xfz} places the Starobinsky model and Higgs inflation firmly within the allowed region and strongly disfavours large-field monomial models with $r\gtrsim 0.05$. Second, the BK18 data break the approximate $n_{s}$--$r$ degeneracy along the attractor trajectory: combined with a precise $n_s$ measurement, the constraint on $r$ directly translates into a constraint on the Starobinsky model as shown in Refs.~\cite{Modak:2022gol,Addazi:2025qra}.

\subsection{Data Combinations}
\label{subsec:datacomb}
We  define the three following datasets for our analysis
\begin{itemize}\setlength\itemsep{3pt}
  \item \textbf{CMB-SPA}: Planck\,2018 low-$\ell$ TT $+$ SPT-3G D1 TnE (lite) $+$ P-ACT (TT+TE+EE $+$ Lensing, \texttt{actplanck\_baseline}) $+$ SPT-3G 2-yr MUSE delensed $\phi\phi$.
  \item \textbf{CMB-SPA+DESI}: SPA $+$ DESI\,DR2 BAO.
  \item \textbf{CMB-SPA+DESI+BK18}: CMB-SPA+DESI $+$ BICEP/\textit{Keck}\,2018.
\end{itemize}
At this point we remark that the optical depth to reionisation is assigned the informative Gaussian prior $\tau_{\rm{reio}}\sim\mathcal{N}\!\left(0.051,\;0.006\right)$
derived from the Planck PR4 large-scale polarisation analysis as adopted in the SPT-3G D1 baseline~\cite{SPT-3G:2025bzu, Planck:2020olo}. This prior is physically motivated on the ground that SPT-3G D1 T\&E power spectrum measurements do not cover large-scale E-mode polarization anisotropies, so they cannot independently constrain $\tau_{\rm{reio}}$ with high precision. As in Ref.~\cite{SPT-3G:2025bzu}, without large-scale E-mode information, there is a degeneracy between $A_s$ and the suppression factor $e^{-2\tau_{\rm{reio}}}$, which can only be broken via gravitational lensing.

The overall ACT calibration parameter is constrained by the external prior
\begin{equation}
  A_{\rm ACT}\sim\mathcal{N}\!\left(1.000,\;0.003\right)\,,
  \label{eq:prior_Aact}
\end{equation}
implemented in the Cobaya likelihood configuration, following the ACT DR6 convention~\cite{AtacamaCosmologyTelescope:2025blo}.

The priors for the seven BK18 foreground parameters $A_{B,\rm dust}$, $A_{B,\rm sync}$, $\alpha_{B,\rm dust}$, $\alpha_{B,\rm sync}$, $\beta_{B,\rm dust}$, $\beta_{B,\rm sync}$ and $\epsilon_{\rm dust,sync}$ are assumed as prescribed in the official BK18 likelihood release~\cite{BICEP:2021xfz}. Note that, the dust spectral index $\beta_{B,\rm dust}$ is assigned a broad flat prior over $[0.80, 2.40]$ rather than the external Gaussian prior used in earlier BICEP/Keck releases. This relaxation is possible because the addition of the BICEP3 220~GHz channel in BK18 provides direct, sufficient in-band frequency leverage to constrain dust spectral behaviour internally. In contrast, the synchrotron spectral index $\beta_{B,\rm sync}$ retains its standard informative Gaussian prior $\mathcal{N}(-3.1, 0.30)$ due to signal-to-noise limitations at the lowest frequencies~\cite{BICEP:2021xfz}.

\section{The Starobinsky Model}
\label{sec:Staro}
In this section we investigate the constraints on baseline Starobinsky model. We implemented the inflationary potential in the Boltzmann code CLASS~\cite{Blas:2011rf,Lesgourgues:2011rg} and interfaced it with Cobaya~\cite{Torrado:2020dgo} for sampling and parameter inferences. Unless otherwise specified, we use flat prior on both the inflationary and $\Lambda$CDM parameters.

\subsection{Model framework and parameter constraints}
\label{subsec: modelframework}

The Starobinsky model~\cite{Starobinsky:1980te} is one of the most celebrated inflationary model, where the Einstein-Hilbert action is modified by an addition of dimension-four quadratic Ricci scalar coupling. The action reads as
\begin{align}
  S_J  =  \int\text{d}^4x \sqrt{-g_J}\: \frac{M_P^2}{2} f(R_J) \; ,
  \label{eq:acJordan}
\end{align}
where $g_J$ denotes the determinant of the mostly plus space-time metric
${g_{\mu\nu}}_J$ and $M_P= (8\pi
G)^{-1/2}$. The  $f(R_J)$ is given as
\begin{align}
  f(R_J) = \left( R_J+ \frac{1}{6 M^2} R_J^2 \right).
\end{align}
For convenience, we study the dynamics of $R^2$ inflation in Einstein frame, where a physical scalar degrees of freedom emerge with  kinetic term having correct sign. To achieve this, we first convert the generic $f(R_J)$ action in Eq.~\eqref{eq:acJordan} into a scalar-tensor theory via a Legendre transformation by introducing an
auxiliary field $\Psi$
\begin{align}
  S_J  =  \int d^4 x \sqrt{-g_J}  \bigg[&\frac{M_{P}^{2}}{2} \left(f(\Psi)
  + \frac{\partial f(\Psi)}{\partial \Psi} (R_J-\Psi)\right)\bigg]\label{eq:actionJ3}.
\end{align}
The Legendre transformation is well defined if  $f(R_J)$ is convex, which translates to $\frac{1}{3 M^2} > 0$ and holds for $M^2 > 0$.
We introduce a physical degree of freedom
\begin{align}
\Theta = \frac{\partial f(\Psi,  \Phi)}{\partial \Psi},\label{eq:theta}
\end{align}
and rewrite the action in Eq.~\eqref{eq:actionJ3} as
\begin{align}
  S_J  =  \int d^4 x \sqrt{-g_J} \bigg[& \frac{M_{P}^{2}}{2} \Theta R_J - U(\Theta)  \bigg],\label{eq:actionJ4}
\end{align}
with,
\begin{align}
\Psi &= 3 M^2 \left(-1 + \Theta\right),\\
U(\Theta) &= \frac{M_{\rm P}^2}{2}\left[\Psi \Theta - f(\Psi)\right]
\end{align}

The Eq.\eqref{eq:acJordan} can now be written in the Einstein frame via Weyl rescaling ${g_{\mu\nu}}_E = \Theta
{g_{\mu\nu}}_J$,
\begin{align}
 S_E =  \int\text{d}^4x \sqrt{-g_E}\:\left[\frac{M_P^2}{2} R_E -\frac{1}{2} g^{\mu\nu}_E
 \left(\nabla_\mu \varphi \nabla_\nu \varphi\right)- V_E(\varphi) \right],\label{eq:acEin}
\end{align}
with the canonical field $\varphi$ and the potential $V_E(\varphi)$ are defined as
\begin{align}
&\varphi = \sqrt{\frac{3}{2}} M_P \ln \Theta ,\label{eq:canofield}\\
& V_E(\varphi) = \frac{1}{\Theta^2} U(\Theta) = \frac{3 M_P^2 M^2}{4}\left(1- e^{-\sqrt{\frac{2}{3}}\frac{\varphi}{M_P}}\right)^2  \label{eq:canopot},\\
&R_J = \Theta \left[R_E + 3  \Box_E{\ln\Theta}-\frac{3 }{2} g^{\mu\nu}_E \partial_\mu \ln\Theta \ \partial_\nu \ln\Theta\right].
  \label{eq:Ricci}
\end{align}
The $\nabla_\mu$ is spacetime covariant derivative which reduces to ordinary partial derivative for scalar field $\varphi$
and $\Box_E = g^{\mu\nu}_E \partial_\mu \partial_\nu$ is the
d’Alembert operator. The $\varphi_*$ i.e., the field value of the inflaton when the pivot scale $k_*$ exits horizon
and the mass parameter $M$ can account for the observed nearly-scale invariant observed CMB power spectrum and spectral index.
The parameter $\varphi_*$ can be traded with $N_*$ i.e., the number of $e$-folds between when the reference mode exit horizon and the
end of inflation.

\begin{figure}[h]
  \centering
  \includegraphics[width = 0.7\textwidth]{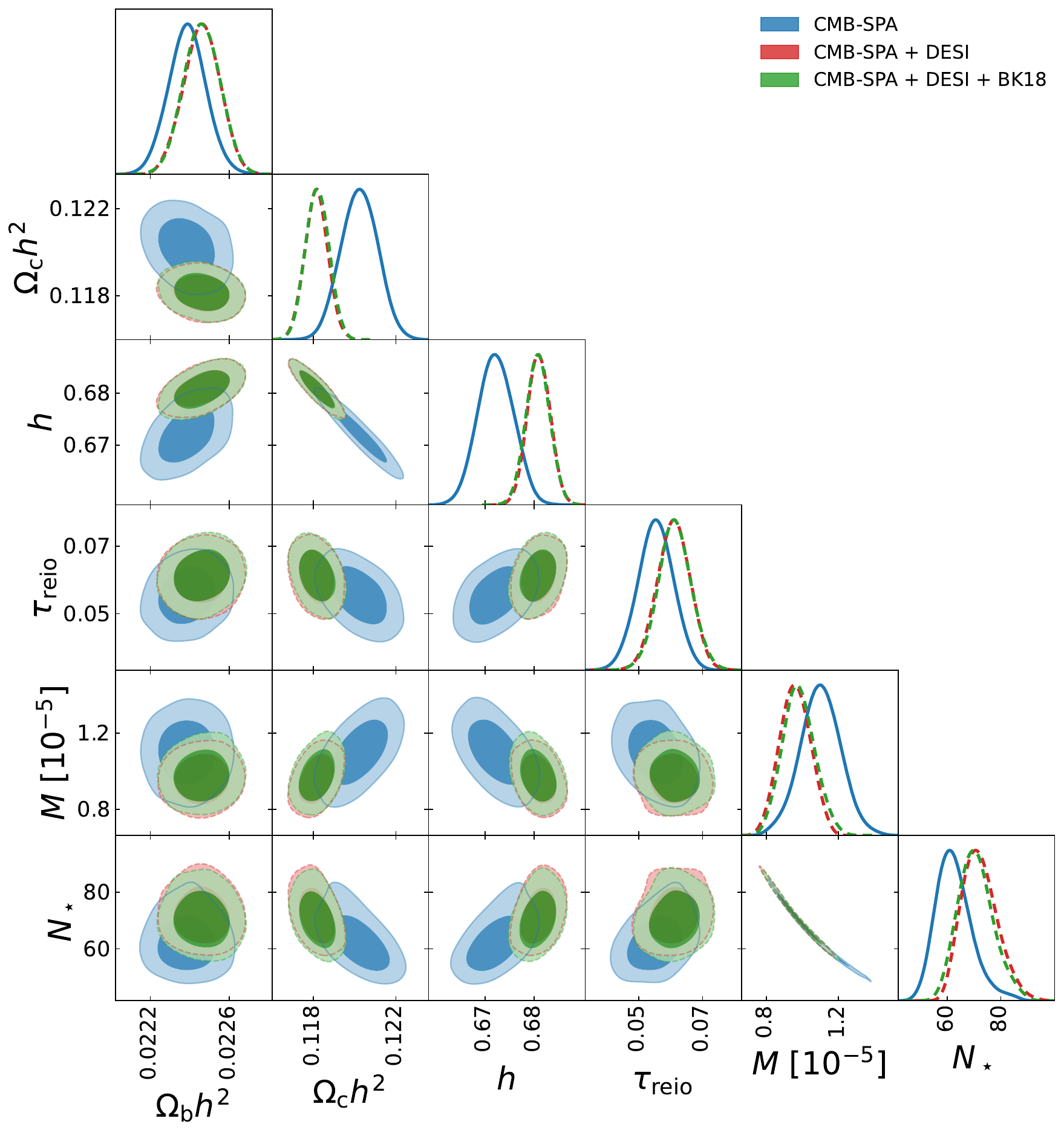}
  \caption{The posterior distributions of the baseline Starobinsky ($R^2$) inflation for the three dataset combinations CMB-SPA (blue), CMB-SPA+DESI (red) and CMB-SPA+DESI+BK18 (green).
  The darker shaded regions denoted 68\% and 95\% CI regions. See text for detailed discussions.}
  \label{fig:staropure}
\end{figure}

The potential $V_E$ in Eq.~\eqref{eq:canopot} is implemented directly in CLASS for global analysis, which solves the equations of motion for the inflaton and the perturbations exactly. The model parameters along with $\Lambda$CDM parameters and, $M$ and $N_*$, with $\ln(\frac{a(t_{\rm end})}{a(t_*)})$, constitute the baseline model parameters
\begin{align}
\{\; \omega_\text{b} , \omega_\text{cdm} , h , \tau_{\rm{reio}}, M , N_* \; \} \; .
\label{eq:model_paras}
\end{align}
sampled in Cobaya with flat prior for all parameters except for $\tau_{\rm{reio}}$ and $N_*$. For the optical depth to reionisation $\tau_{\rm{reio}}$ assumed to have an informative Gaussian prior $\tau_{\rm{reio}}\sim\mathcal{N}\!\left(0.051,\;0.006\right)$ which is adopted from the Planck PR4 large scale polarisation analysis as adopted in the SPT-3G D1 baseline~\cite{SPT-3G:2025bzu, Planck:2020olo}. This prior is physically motivated on the ground that SPT-3G D1 T\&E power spectrum measurements do not cover large-scale E-mode polarization anisotropies, so they cannot independently constrain $\tau_{\rm{reio}}$ with high precision. Here, without the large scale E-mode information, there is a degeneracy between $A_s$ and the suppression factor $e^{-2\tau_{\rm{reio}}}$, which can only be broken via gravitational lensing~\cite{SPT-3G:2025bzu}. We take the Gaussian prior for $N_* \in \mathcal{N}(60,20)$; we shall return to the prior dependence on $N_*$ shortly. We have also set $\varphi_{\rm end}\approx 0$ for simplicity in CLASS while sampling for the model. The chains are assumed to have converged for the baseline Starobinsky model when the Gelman-Rubin criterion $\hat{R}-1\lesssim0.03$. We consider at least six chains with each chain containing at least $1\times 10^{5}$ steps for any particular inference.

The marginalised posterior distributions for the parameters are plotted in Fig.~\ref{fig:staropure} for the three combinations CMB-SPA, CMB-SPA+DESI and CMB-SPA+DESI+BK18. The corresponding   mean values, 68\% credible interval (CI) and best-fit values for the parameters are summarized in Table~\ref{tab:staropure}. The best-fit values are obtained using the BOBYQA minimizer implemented in Cobaya~\cite{Torrado:2020dgo}. We find that, for the combinations with BAO data, $N_*$ is pushed towards higher values compared to CMB-only combination CMB-SPA.
This trend is consistent with $N_*$ derived from the marginalized $n_s$ under slow-roll approximation. However, a comparison can be made now between the constraint on $N_*$ that we obtain and those inferred from the mean and uncertainty of $n_s$ in Ref.~\cite{SPT-3G:2025bzu}, using the slow-roll relation $N_* \simeq 2/\left(1-n_s\right)$, as commonly adopted in the literature. E.g., the CMB-S4+DESI constraints $n_s = 0.9728 \pm 0.0027$~\cite{SPT-3G:2025bzu} translates to $N_* \approx 73.53^{+8.10}_{-6.64}$. In contrast, we find the mean $N_*=72.08^{+5.60}_{-7.55}$ and best-fit 71.06.  The Ref.~\cite{SPT-3G:2025bzu} reported constraint $n_s=0.9684\pm0.003$ for the CMB-SPA combination which translates to $N_* \approx 63.29^{+7.63}_{-5.49}$. This should be compared to the constraint $N_*=62.53^{+5.25}_{-7.62}$ and best-fit value 62.82 in Table~\ref{tab:staropure}. By simply symmetrizing the 68\% errors the constraints on $N_*$ summarized in Table~\ref{tab:staropure} are a bit stronger than the corresponding ones derived from $n_s$ using slow-roll relation.
We find the CMB-SPA+DESI+BK18 constraint as $N_*=70.66^{+5.79}_{-7.35}$ with best-fit 69.94.

\begin{table}[htbp!]
\centering
\renewcommand{\arraystretch}{1.25}
\resizebox{\columnwidth}{!}{%
\begin{tabular}{|l|c|c|c|c|c|c|}
\hline
& \multicolumn{2}{c|}{\textbf{CMB-SPA}}
& \multicolumn{2}{c|}{\textbf{CMB-SPA+DESI}}
& \multicolumn{2}{c|}{\textbf{CMB-SPA+DESI+BK18}} \\[1.2ex]
\cline{2-7}
\textbf{Sampled} &
\textbf{Mean $\pm$ 68\% CI} & \textbf{Best-fit} &
\textbf{Mean $\pm$ 68\% CI} & \textbf{Best-fit} &
\textbf{Mean $\pm$ 68\% CI} & \textbf{Best-fit} \\[0.8ex]
\hline
$h$ & $0.6722^{+0.0037}_{-0.0036}$ & 0.6722 & $0.6809 \pm 0.0023$ & 0.6823 & $0.6807 \pm 0.0024$ & 0.6820 \\
$\Omega_b h^2$ & $0.022387^{+0.000094}_{-0.000095}$ & 0.022419 & $0.022461^{+0.000092}_{-0.000091}$ & 0.022505 & $0.022460 \pm 0.000093$ & 0.022463 \\
$\Omega_c h^2$ & $0.12023 \pm 0.00090$ & 0.12047 & $0.11814 \pm 0.00056$ & 0.11788 & $0.11818 \pm 0.00057$ & 0.11786 \\
$\tau_{\rm{reio}}$ & $0.05540^{+0.00555}_{-0.00554}$ & 0.05478 & $0.06099^{+0.00517}_{-0.00510}$ & 0.06292 & $0.06130^{+0.00510}_{-0.00498}$ & 0.06077 \\
$M\ [10^{-5}]$ & $1.100^{+0.112}_{-0.110}$ & 1.082 & $0.960^{+0.087}_{-0.086}$ & 0.967 & $0.980^{+0.084}_{-0.094}$ & 0.981 \\
$N_*$ & $62.53^{+5.25}_{-7.62}$ & 62.82 & $72.08^{+5.60}_{-7.55}$ & 71.06 & $70.66^{+5.79}_{-7.35}$ & 69.94 \\
\hline
\end{tabular}}
\caption{The parameter constraints and respective best-fit values of the baseline Starobinsky model parameters. Here we assumed the Gaussian prior $\mathcal{N}(60,20)$ for $N_*$.}
\label{tab:staropure}
\end{table}
We now turn our attention to strong correlation between $N_*$ and $M$ and the prior choice on $N_*$. As in Ref.~\cite{Modak:2022gol}, we find the chains do not converge if flat prior is chosen for $N_*$. This is solely the reason we chose Gaussian prior for $N_*$. Further, the marginalized $N_*$ and $M$ show dependence on the chosen width of the Gaussian prior. This can be understood from how these parameters are constrained by $n_s$ and the scalar amplitude $A_s$. In the slow-roll approximation the $n_s$ is directly related to $N_*$, while $A_s$ constrains the combination $M$ and  $N_*$. This also true beyond the slow-roll analysis that we adopted here. Furthermore, the $r$ is also dependent on $N_*$. Therefore we also find strong degeneracy between $M$ and $N_*$. As a result, the marginalized constraints on $M$ and $N_*$ still carry some dependence on the chosen prior for $N_*$ as found in Fig.~\ref{fig:staropure}. We have checked this by varying the width of the Gaussian prior on $N_*$, and find that the posterior mainly broadens along the degeneracy direction while the preferred region stays essentially unchanged. This simply reflects the underlying parameter degeneracy and should be seen as a feature of the phenomenological setup rather than a direct observational constraint. We further find that the posterior is stable for broad Gaussian priors on $N_*$ with $\sigma \sim 15$--$20$, whereas it starts to change for narrower choices such as $\sigma\lesssim 10$. This indicates that the narrow prior is actually more informative than what the likelihood can constrain along the $N_*$ direction, and it therefore cuts into the allowed degeneracy between $M$ and $N_*$. On the other hand, once the prior is sufficiently broad, the posterior becomes essentially prior-independent and is fully controlled by the degeneracy structure of the likelihood in the $(M,N_*)$ plane.

\begin{figure}[h]
  \centering
 \includegraphics[width = \textwidth]{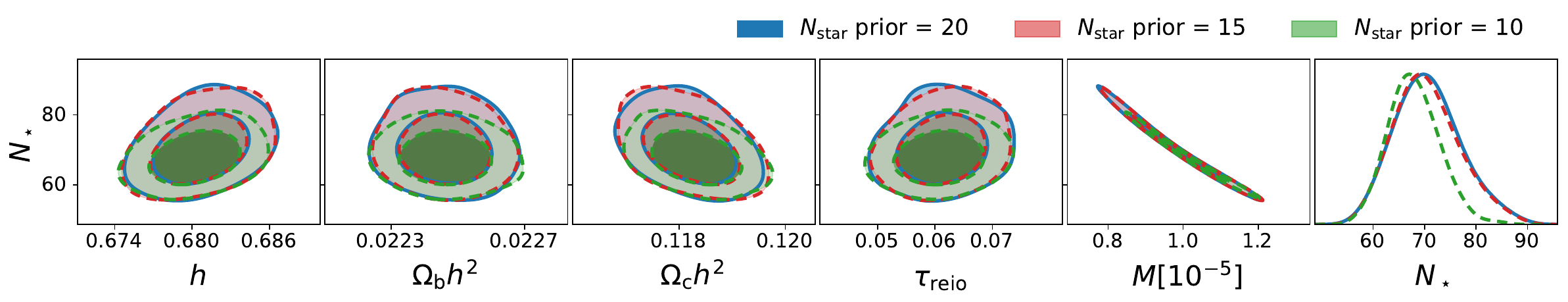}
  \caption{Dependence of the Gaussian $N_*$ prior with width $\sigma=10, 15,$ and $20$ on the posteriors for the CMB-SPA+DESI+BK18 combination.}
  \label{fig:nstar_priors}
\end{figure}

Consequently, we take a conservative prior $N_* \in \mathcal{N}(60,20)$ and choose a relative loose convergence criteria $\hat{R}-1\lesssim0.03$ to balance heavy computational cost.
The effect of Gaussian $N_*$ prior with $\sigma=10, 15,$ and $20$ is shown in Fig.~\ref{fig:nstar_priors} with the corresponding parameter constraints are given in Table~\ref{tab:staropriors} for CMB-SPA+DESI+BK18 combination as illustration. The standard $\Lambda$CDM parameters barely change across the different priors. On the other hand, $M$ and $N_*$ clearly depend on the prior width because of their strong degeneracy. With a narrow prior ($\sigma=10$), we get the constraints on $N_* = 67.93^{+4.58}_{-5.44}$ and $M = 1.014^{+0.069}_{-0.076} \times 10^{-5}$. When we widen the prior to $\sigma=15$ , $N_*$ increases to $70.42^{+5.60}_{-7.40}$ while $M$ drops slightly to $0.983^{+0.089}_{-0.090} \times 10^{-5}$. Intriguingly, for $\sigma=20$, the constraints practically remain unchanged, with $N_*$ mean$\pm68$\% CI reads as $70.66^{+5.79}_{-7.35}$ and while $M$ slightly changes to $0.980^{+0.084}_{-0.094} \times 10^{-5}$. This justifies the choice of $\sigma=20$ for Fig.~\ref{fig:staropure} and Table~\ref{tab:staropure}.

\begin{table}[htbp!]
\centering
\renewcommand{\arraystretch}{1.25}
\resizebox{\textwidth}{!}{%
\begin{tabular}{|l|c|c|c|c|c|c|}
\hline
\multirow{2}{*}{\textbf{Sampled}} & \multicolumn{2}{c|}{\textbf{Prior: $\sigma=10$}}
& \multicolumn{2}{c|}{\textbf{Prior: $\sigma=15$}}
& \multicolumn{2}{c|}{\textbf{Prior: $\sigma=20$}} \\[1.2ex]
\cline{2-7}
& \textbf{Mean $\pm$ 68\% CI} & \textbf{Best-fit} &
\textbf{Mean $\pm$ 68\% CI} & \textbf{Best-fit} &
\textbf{Mean $\pm$ 68\% CI} & \textbf{Best-fit} \\[0.8ex]
\hline
$h$ & $0.6803 \pm 0.0023$ & 0.6815 & $0.6806 \pm 0.0024$ & 0.6815 & $0.6807 \pm 0.0024$ & 0.6820 \\
$\Omega_b h^2$ & $0.022463^{+0.000090}_{-0.000091}$ & 0.022513 & $0.022460^{+0.000086}_{-0.000087}$ & 0.022463 & $0.022460 \pm 0.000093$ & 0.022463 \\
$\Omega_c h^2$ & $0.11831^{+0.00055}_{-0.00056}$ & 0.11808 & $0.11822^{+0.00058}_{-0.00061}$ & 0.11795 & $0.11818 \pm 0.00057$ & 0.11786 \\
$\tau_{\rm{reio}}$ & $0.06082^{+0.00506}_{-0.00513}$ & 0.06209 & $0.06128^{+0.00488}_{-0.00504}$ & 0.06314 & $0.06130^{+0.00510}_{-0.00498}$ & 0.06077 \\
$M\ [10^{-5}]$ & $1.014^{+0.069}_{-0.076}$ & 1.016 & $0.983^{+0.089}_{-0.090}$ & 0.983 & $0.980^{+0.084}_{-0.094}$ & 0.981 \\
$N_*$ & $67.93^{+4.58}_{-5.44}$ & 67.55 & $70.42^{+5.60}_{-7.40}$ & 69.85 & $70.66^{+5.79}_{-7.35}$ & 69.94 \\
\hline
\end{tabular}
}
\caption{The Gaussian $N_*$ prior with widths of $\sigma=10, 15,$ and $20$ dependence on the parameter constraints and the best-fit values for the CMB-SPA+DESI+BK18 combination.}
\label{tab:staropriors}
\end{table}

\begin{figure}[h]
  \centering
 \includegraphics[width = 0.6 \textwidth]{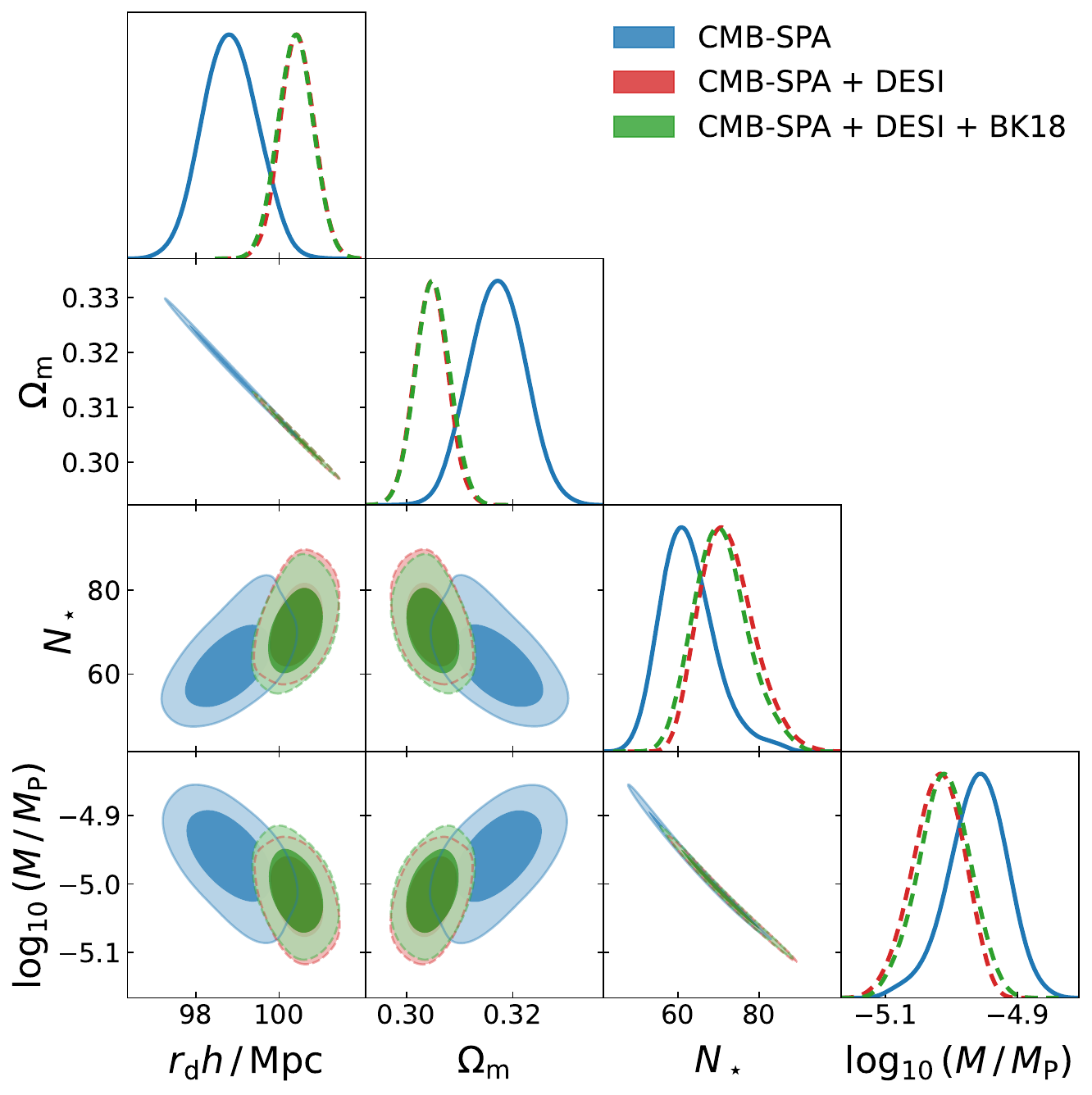}
  \caption{The correlation of $M$ and $N_*$ with the BAO parameter $r_d h$ for three data combinations for comparison.}
  \label{fig:rdplot}
\end{figure}

The CMB-BAO tension, which is discussed above, is also visible in the correlation between $N_*$ and $M$ with $\Omega_c h^2$ and $h$, as can be seen in Fig.~\ref{fig:staropure}. We find that for combinations with DESI BAO, i.e. CMB-SPA+DESI  and  CMB-SPA+DESI+BK18, the $N_*$ mean pushed towards higher values in contrast to CMB-SPA. An opposite trend is found for $M$, which shifted towards lower values for combinations with DESI BAO. This is consistent with the finding of Ref.~\cite{Ferreira:2025lrd} where a shift of the $n_s$ to higher value is attributed to the correlation of $n_s$ and BAO parameter $r_d h$ between CMB-only and DESI BAO data. For illustration, we plot the correlation of the $M$ and $N_*$ with $r_d h$ in Fig.~\ref{fig:rdplot}. We find the same correlation is reflected here.

\subsection{Reheating temperature}
\label{subsec:reheating}

Finally, one may construct the dependence between of the model parameter $M$ with the reheating temperature $T_{\rm re}$ under slow-roll approximation. The CMB pivot scale $k_{\rm ref}/a_0 = 0.05~{\rm Mpc}^{-1}$ (with $a_0$ being scale factor today), is related to the reference mode $k_*$ that exit horizon at $N_*$ is related through \begin{align}
k_{\rm ref} =  k_* = a(t_*) H(t_*)= \frac{a(t_*)}{a(t_{\rm end})} \frac{a(t_{\rm end})}{a(t_{\rm re})} \frac{a(t_{\rm re})}{a_0} a_0 H(t_*), \label{eq:kefrel}
\end{align}
where $a(t_{\rm re})$ denotes the scale factor at the end of the reheating. Here $k_*$ is mode that exits horizon at $N_*$. We assume that the radiation dominated epoch begins immediately after reheating.
We can now reexpress Eq.~\eqref{eq:kefrel} as
\begin{align}
N_* = \ln\Biggl[\frac{H_*}{k_{\rm ref}/a_0} \frac{T_0}{T_{\rm re}} \frac{g_0^{1/3}}{g_{\rm re}^{1/3}}\Biggr] + N_{\rm re}, \label{eq:nstar}
\end{align}
where $N_{\rm re}=\ln\left(\frac{a(t_{\rm end})}{a(t_{\rm re})}\right)$ and we have utilized
\begin{align}
\rho_{\mathrm{inf}}\bigr|_{N=N_{\rm re}}\equiv\rho_{\rm re}=\frac{g_{\rm re} \pi^2}{30} \; T_{\rm re}^4,\label{eq:prehettemp}
\end{align}
with $T_0=2.7 K$  is the temperature today and $g_0= 43/11$ and $g_{\rm{re}}=106.75$ are the relativistic degrees of freedom today and end of reheating. Under slow-roll approximation
\begin{align}
N_* &= \frac{1}{M_{\rm P}^2}
\int_{\varphi_{\rm end}}^{\varphi_*}
\frac{V_E}{V_{E,\varphi}}\, d\varphi,
\end{align}
which translates  to
\begin{equation}
H_* \simeq \frac{M}{2}
\left(1-\frac{3}{4N_*} \right).\label{eq:slow}
\end{equation}

\begin{figure}[h]
  \centering
  \includegraphics[width = 0.32\textwidth]{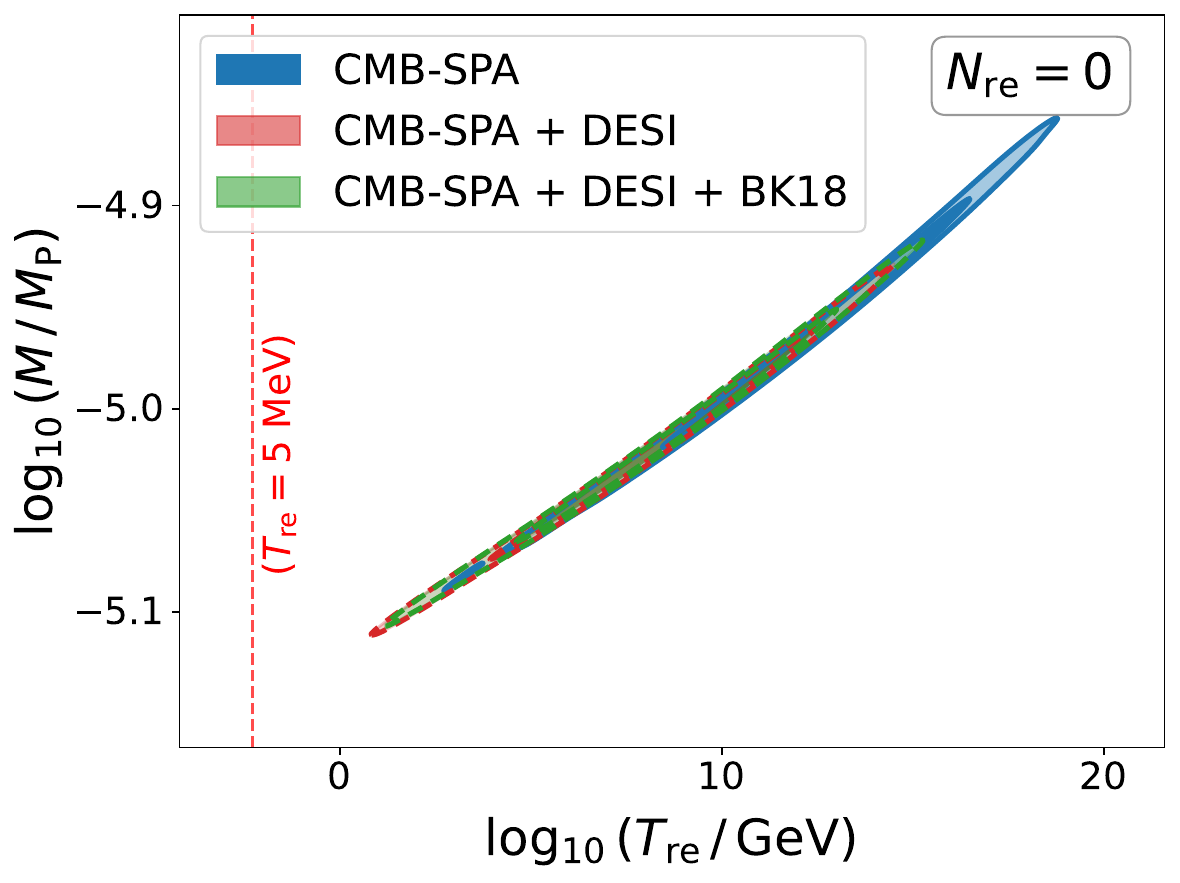}
  \includegraphics[width = 0.32\textwidth]{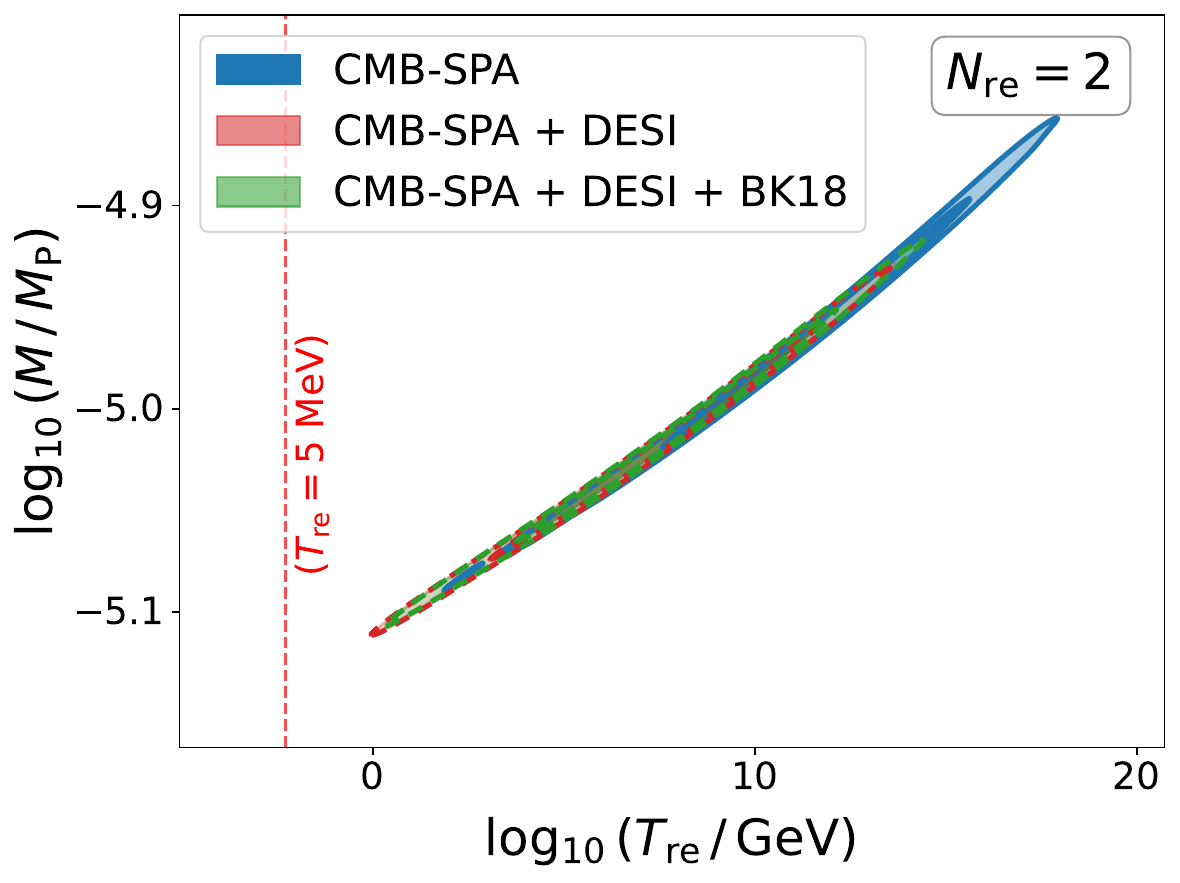}
  \includegraphics[width = 0.32\textwidth]{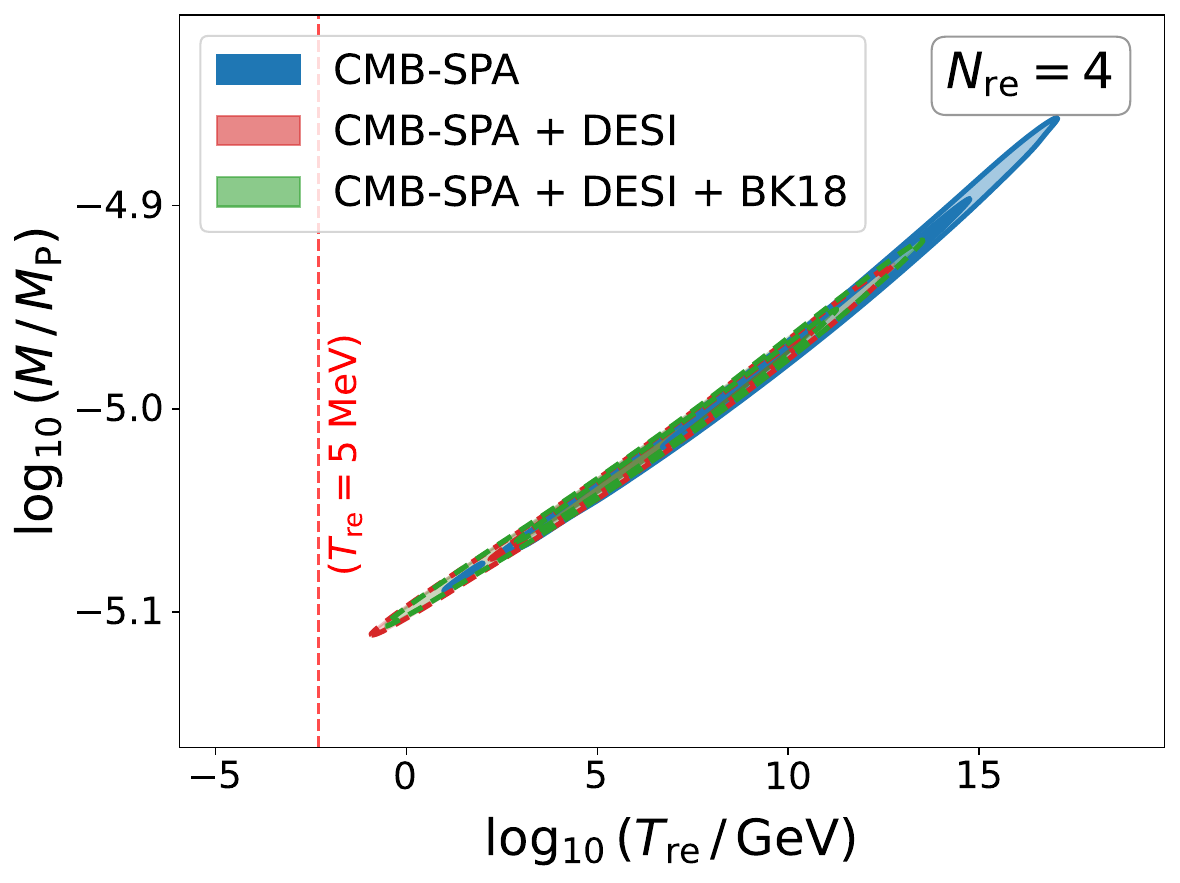}
  \caption{The $M$ vs $T_{\rm re}$ for pure Starobinsky model. See text for detailed discussions.}
  \label{fig:TrevsM}
\end{figure}

Substituting Eq.~\eqref{eq:slow} into Eq.~\eqref{eq:nstar}, and utilizing the posterior of $M$ and $N_*$, we derive the constraints in the $T_{\rm re}$ vs $M$ plane, shown in Fig.~\ref{fig:TrevsM} for fixed $N_{\rm re}=0$, 2 and 4. Because $N_*$ and $T_{\rm re}$ are directly connected through Eqs.~\eqref{eq:nstar} and \eqref{eq:slow}, the strong correlation between $M$ and $N_*$ is correspondingly manifested as a strong correlation between $M$ and $T_{\rm re}$ for a given value of $N_{\rm re}$; in other words the degeneracy between $M$ and $N_*$ can be broken for a fixed $T_{\rm re}$.  We remark that, given $N_{\rm re}$, one has the freedom to decide the particle content and the exact reheating mechanism needed to satisfy the data. In this way, we differentiate which parameters CMB actually constrains and, the uncertainties regarding the post-inflationary reheating history is left on a single parameter  $N_{\rm re}$.

The derived constraints and the best-fit values of $T_{\rm re}$ are summarized in Table~\ref{tab:Tre_log10_values} for different values of $N_{\rm re}$. For $N_{\rm re}=0$, corresponding to instantaneous reheating, the best-fit reheating temperature is $T_{\rm re}\approx 1.64\times10^{12}$ GeV for CMB-SPA, which become $\sim 3.86\times 10^{8}$ GeV for CMB-SPA+DESI and $\sim 1.21\times 10^{9}$ GeV for the CMB-SPA+DESI+BK18 datasets. For $N_{\rm re}=2$ (4), the best-fit values of $T_{\rm re}$ are $\sim 2.22\times 10^{11}$ ($3.01\times 10^{10}$) GeV for CMB-SPA, $5.22\times 10^{7}$ ($7.07\times 10^{6}$) GeV for CMB-SPA+DESI, and $1.63\times 10^{8}$ ($2.21\times 10^{7}$) GeV for CMB-SPA+DESI+BK18. While the best-fit values of $T_{\rm re}$ are quite indicative, however, it should be noted that the constraints on $T_{\rm re}$ with different data combinations still allows the possibility of different exotic reheating mechanism along with the standard ones. Our analysis yields somewhat different constraints on $T_{\rm re}$ than Ref.~\cite{Zharov:2025zjg} (see also Refs.~\cite{Drees:2025ngb,Liu:2025qca,Haque:2025uis}), primarily because the post-inflationary evolution is parametrized differently. Ref.~\cite{Zharov:2025zjg} samples the Starobinsky model parameter together with the reheating parameter $R_{\rm reh}$ (while fixing the baseline $\Lambda$CDM parameters), where $R_{\rm reh}$ is defined using the effective relation $\rho \propto a^{-3(1+\bar{w}_{\rm reh})}$. In contrast, in our analysis we combined all uncertainties in post-inflationary reheating including the duration of thermalization within $N_{\rm re}\equiv\ln(a_{\rm re}/a_{\rm end})$. Therefore $N_{\rm re}$ is a purely kinematic quantity, and hence, we do not need to assume any averaged over equation of state. Consequently, the inflationary parameter inference is performed independently, with the post-inflationary uncertainty isolated to the assumed reheating duration $N_{\rm re}$. We therefore leave $N_{\rm re}$ as a free phenomenological parameter, allowing it to be specified according to the particle content and reheating mechanism of the underlying model.

\begin{table}[h]
\centering
\renewcommand{\arraystretch}{1.25}
\begin{tabular}{|l|c|c|c|}
\hline
\multirow{2}{*}{\textbf{$N_{\rm re}$}} & \textbf{CMB-SPA} & \textbf{CMB-SPA+DESI} & \textbf{CMB-SPA+DESI+BK18} \\[1.2ex]
\cline{2-4}
& \textbf{Mean $\pm$ 68\% CI (Best-fit)} & \textbf{Mean $\pm$ 68\% CI (Best-fit)} & \textbf{Mean $\pm$ 68\% CI (Best-fit)} \\[0.8ex]
\hline
0 & $12.35_{-2.32}^{+3.36}$ ($12.22$) & $8.14_{-2.47}^{+3.32}$ ($8.59$) & $8.77_{-2.55}^{+3.23}$ ($9.08$) \\
2 & $11.48_{-2.32}^{+3.36}$ ($11.35$) & $7.27_{-2.47}^{+3.32}$ ($7.72$) & $7.90_{-2.55}^{+3.23}$ ($8.21$) \\
4 & $10.61_{-2.32}^{+3.36}$ ($10.48$) & $6.40_{-2.47}^{+3.32}$ ($6.85$) & $7.03_{-2.55}^{+3.23}$ ($7.34$) \\
\hline
\end{tabular}
\caption{The parameter constraints and best-fit values for $\log_{10}\left(T_{\rm re}/\text{GeV}\right)$ for the baseline Starobinsky model across the CMB-SPA, CMB-SPA+DESI and CMB-SPA+DESI+BK18 combinations for fixed $N_{\rm re}=0, 2,$ and $4$.}
\label{tab:Tre_log10_values}
\end{table}

The reheating temperature predicted in the baseline Starobinsky model with Standard Model (SM) particle content via minimal gravitational reheating is $\sim 10^8$--$10^9$ GeV (see e.g. Refs.~\cite{Dorsch:2024nan,Choi:2019osi,Gorbunov:2010bn}). The best-fit $T_{\rm re}$ values from CMB-SPA for $N_{\rm re}=0,2,$ and $4$ are higher, however, gravitational reheating via SM remains consistent within the $\sim68\%$ CI. In contrast, the best-fit values from the CMB-SPA+DESI and CMB-SPA+DESI+BK18 combinations are fully consistent with minimal gravitational reheating. These results can be straightforwardly extended to other Starobinsky-like scenarios, such as $R^2$-Higgs inflation, provided inflation occurs in the $R^2$-like regime driven by the field $\varphi$. Note that, the best-fit value of the reheating temperature for the single-field like-regime for $R^2$-Higgs inflation range between $\sim10^{13}$--$10^{14}$ GeV for ballpark  $N_{\rm re}=2$--3~\cite{Cado:2023zbm}. This illustrates that the $R^2$-Higgs inflation is still a best-fit model to CMB data.

\section{Starobinsky Inflation with $R^3$ modification}
\label{sec:R3mod}

We extend the original Starobinsky model by a $R^3$-curvature term,
\begin{align}
  f(R_J) = M_P^2 \left(R_J + \frac{1}{6 M^2} R_J^2 + \frac{c}{36 M^4} R_J^3\right) \; ,
  \label{eq:R3}
\end{align}
where $c$ is a dimensionless coefficient, which can be generated by
quantum corrections. Following the prescription as in the case of baseline Starobinsky model, we find two solutions for $\Psi$
\begin{align}
    \Psi=
\begin{cases}
    \dfrac{+ 2 M^2}{c} \left[\sqrt{1+ 3 c (\Theta-1)}-1\right]& \text{for positive branch}\\
    \dfrac{-2 M^2}{c} \left[\sqrt{1+ 3 c (\Theta-1)}+1\right]& \text{for negative branch}, \ \label{eq:svarphi}
\end{cases}
\end{align}
with convexity condition $\Psi > -M^2/(2c)$.

The potential can be expressed as
\begin{align}
 V_E = \frac{1}{\Theta^2} U(\Theta),
\end{align}
with the canonical field is defined as $\varphi = \sqrt{\frac{3}{2}} M_P \ln \Theta$.

Let us now briefly discuss about the positive and negative branch of the solutions of $\Psi$ as in Eq.~\eqref{eq:svarphi}. For the negative branch solutions $V_E$ remains negative practically in the entire $R^2$-like regime i.e., when $M$ is in the reference parameter ranges of the baseline Starobinsky inflation, for both $c >0$ and $c < 0$. However, for certain parameter ranges, we find that $V_E$ can indeed become positive with satisfying the convexity condition but the duration of slow-roll of the inflaton is not sufficient to sustain inflation within our scan range. Hence we neglect the negative branch for finite computation resources. For the positive branch, for $c > 0$ as well as for $c < 0$, the convexity can be maintained and inflaton field can slow-roll sufficiently long enough to sustain inflation and can account for observed high $n_s$~\cite{Gialamas:2025ofz,Kim:2025dyi,Addazi:2025qra,Modak:2025bjv,Park:2025upd}. Thus in our analysis we focus on the positive branch solution in Eq.~\ref{eq:svarphi}. For the positive branch, the potential becomes
\begin{align}
 V_E &=  e^{-2\sqrt{\frac{2}{3}}\frac{\varphi}{M_P}} \frac{M^2 M_{\rm P}^2}{9 c^2}\bigg(1-\sqrt{1+3 c \left( e^{\sqrt{\frac{2}{3}}\frac{\varphi}{M_P}}-1\right)}\bigg)^2
 \bigg(1+2\sqrt{1+3 c \left(e^{\sqrt{\frac{2}{3}}\frac{\varphi}{M_P}}-1\right)}\bigg).\label{eq:potR3}
\end{align}
For $c > 0$ in the positive branch, the best-fit value of the observed high $n_s$ can be satisfied but required $N_*$ preferably be $ > 60$. In such case the potential remains well behaved for $\varphi \gg M_{\rm P}$ and the $c > 0$ the convexity condition is satisfied automatically~\cite{Gialamas:2025ofz,Kim:2025dyi,Addazi:2025qra,Park:2025upd}. On the contrary, for $c <0$, the convexity condition puts a cut-off on $\varphi$~\cite{Gialamas:2025ofz,Addazi:2025qra} but required $N_*$ could still remain below 60 $e$-folding~\cite{Gialamas:2025ofz,Kim:2025dyi,Addazi:2025qra,Park:2025upd}. While  $\varphi_*$ required to account for observed $n_s$ remains below this cut-off for $c < 0$, there still remains subtlety on the behavior of the potential near this cut-off and, how the other higher dimensional terms beyond $R^3$ contribute.

\begin{figure}[htbp!]
  \centering
    \includegraphics[width=0.47\textwidth]{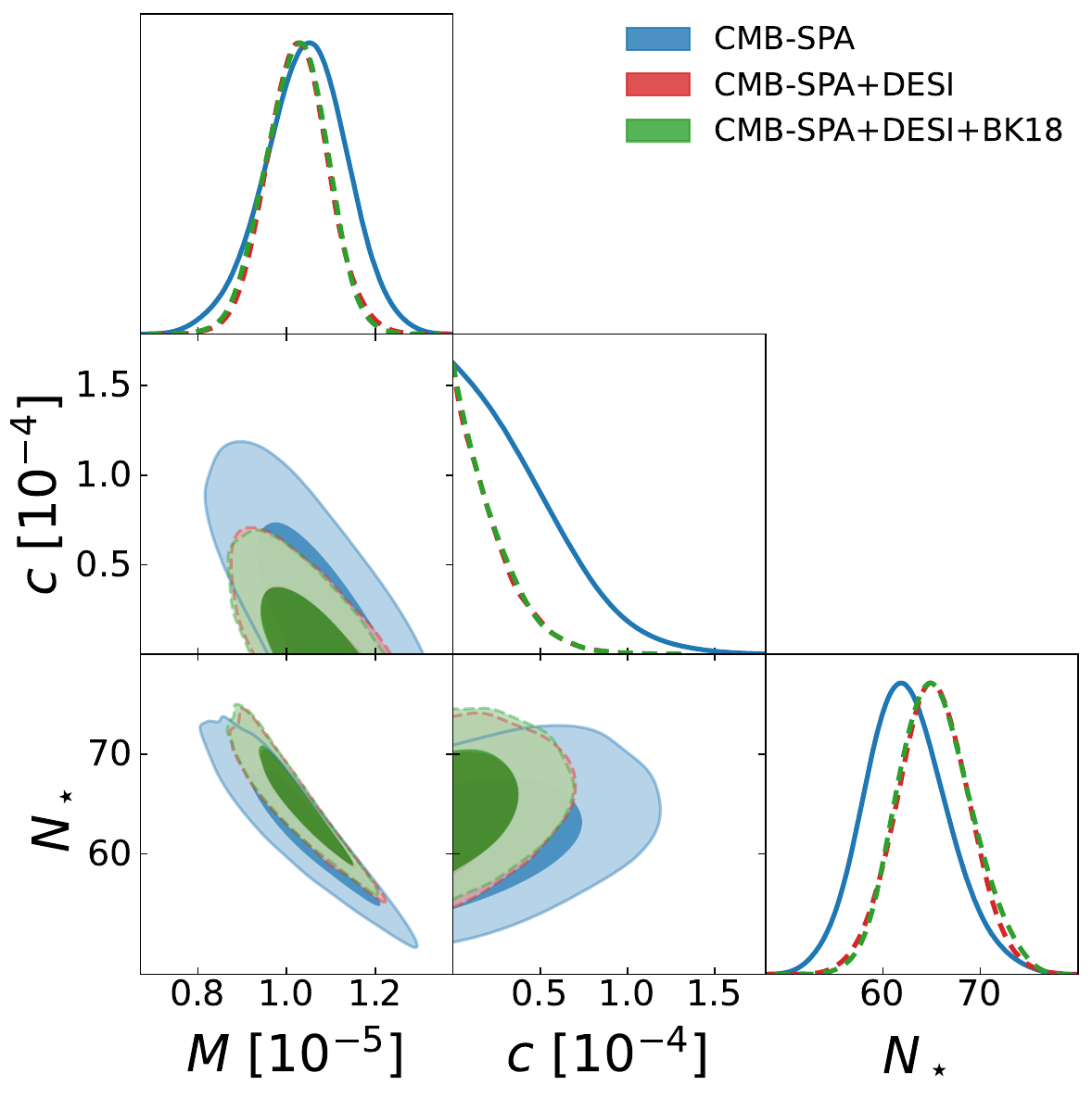}
  \includegraphics[width=0.47\textwidth]{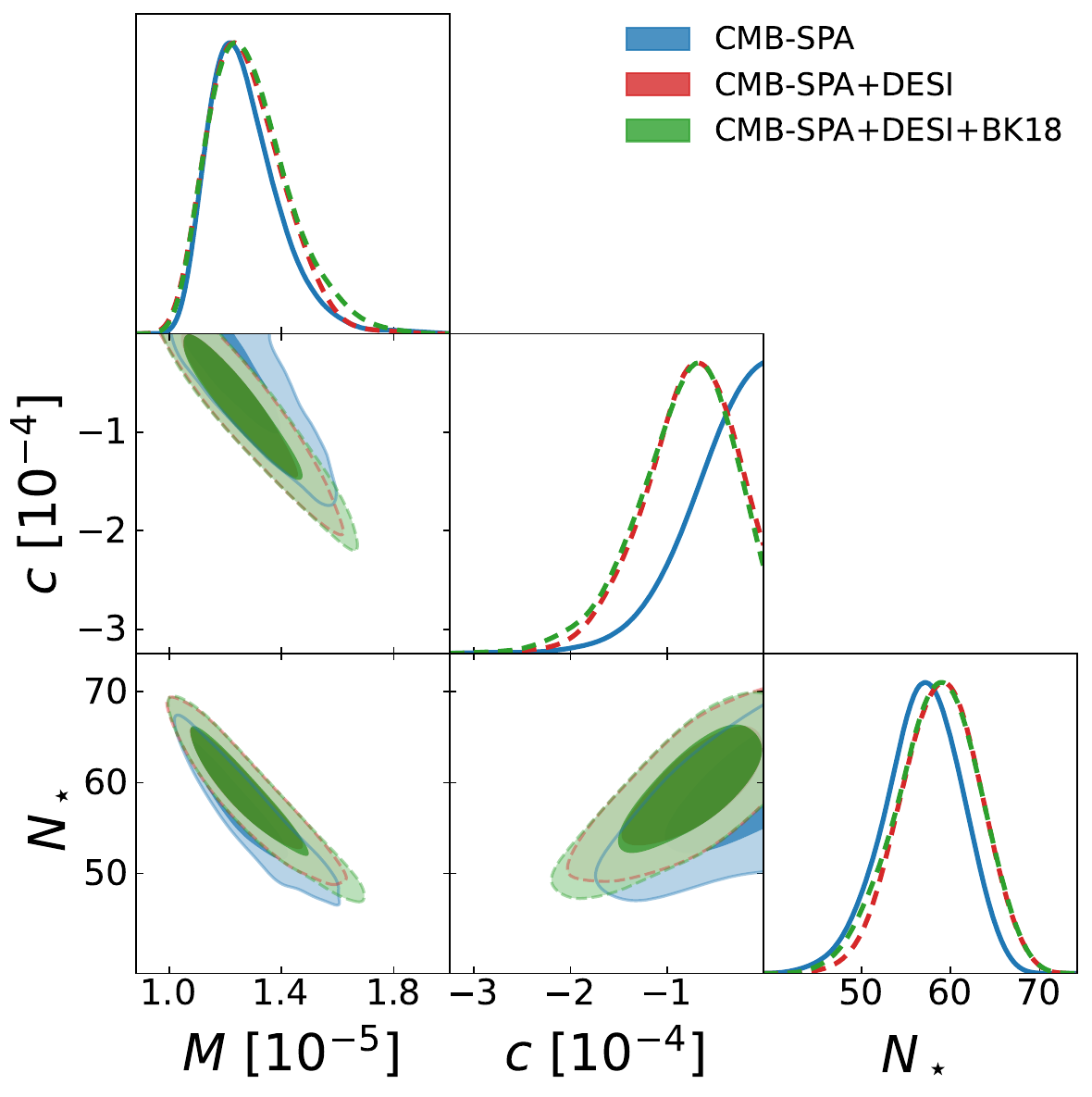}
  \caption{The marginalised posteriors across all three dataset combinations for $R^3$ modified Starobinsky inflation for $c > 0$ (left panel) and $c <0$ (right panel) respectively.}
  \label{fig:r3}
\end{figure}

In Fig.~\ref{fig:r3} and Table~\ref{tab:r3}, we provide the corresponding posterior distributions and  parameter constraints of the model for CMB-SPA, CMB-SPA+DESI and CMB-SPA+DESI+BK18 combinations. We remark that, while generating Fig.~\ref{fig:r3} and Table~\ref{tab:r3}, we sampled the $c >0$ and $c < 0$ separately. Further, as the baseline Starobinsky model suffers from convergence due to parameter degeneracy, having one additional parameter $c$, the posterior of the $R^3$ modified model also suffers from such degeneracy. We adopt a Gaussian prior for $N_*$, centered about 60 with $\sigma_{N_*}=5$. A flat prior is used for $M$ over the range $M \in [5.0\times10^{-6},\,2.0\times10^{-5}]$. For $c$, we adopt a flat prior, varied over $c \in [0,\,0.005]$ for the positive branch and $c \in [-0.005,\,0]$ for the negative branch.

\begin{figure}[h]
  \centering
 \includegraphics[width = 0.47 \textwidth]{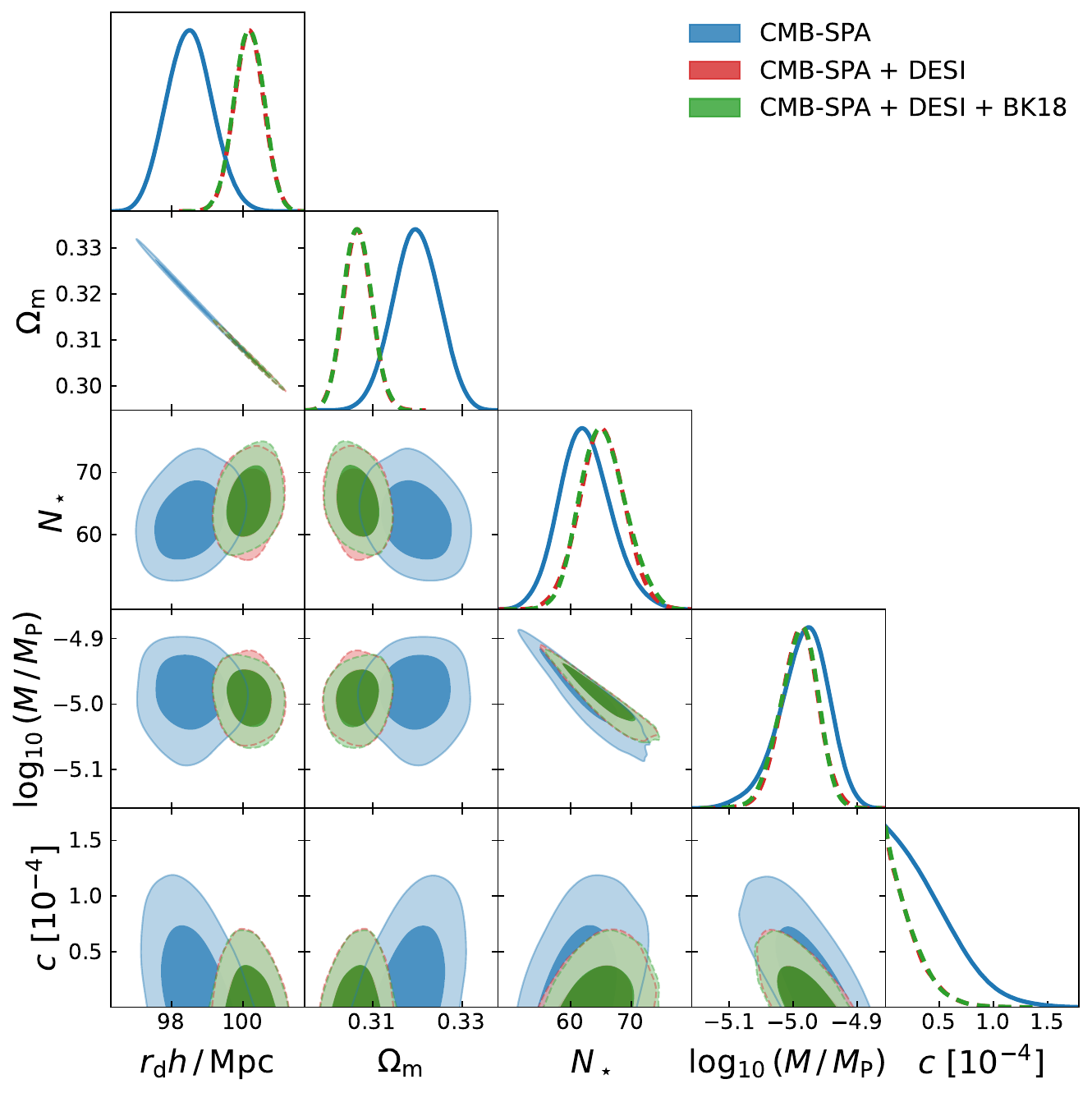}
\includegraphics[width = 0.47 \textwidth]{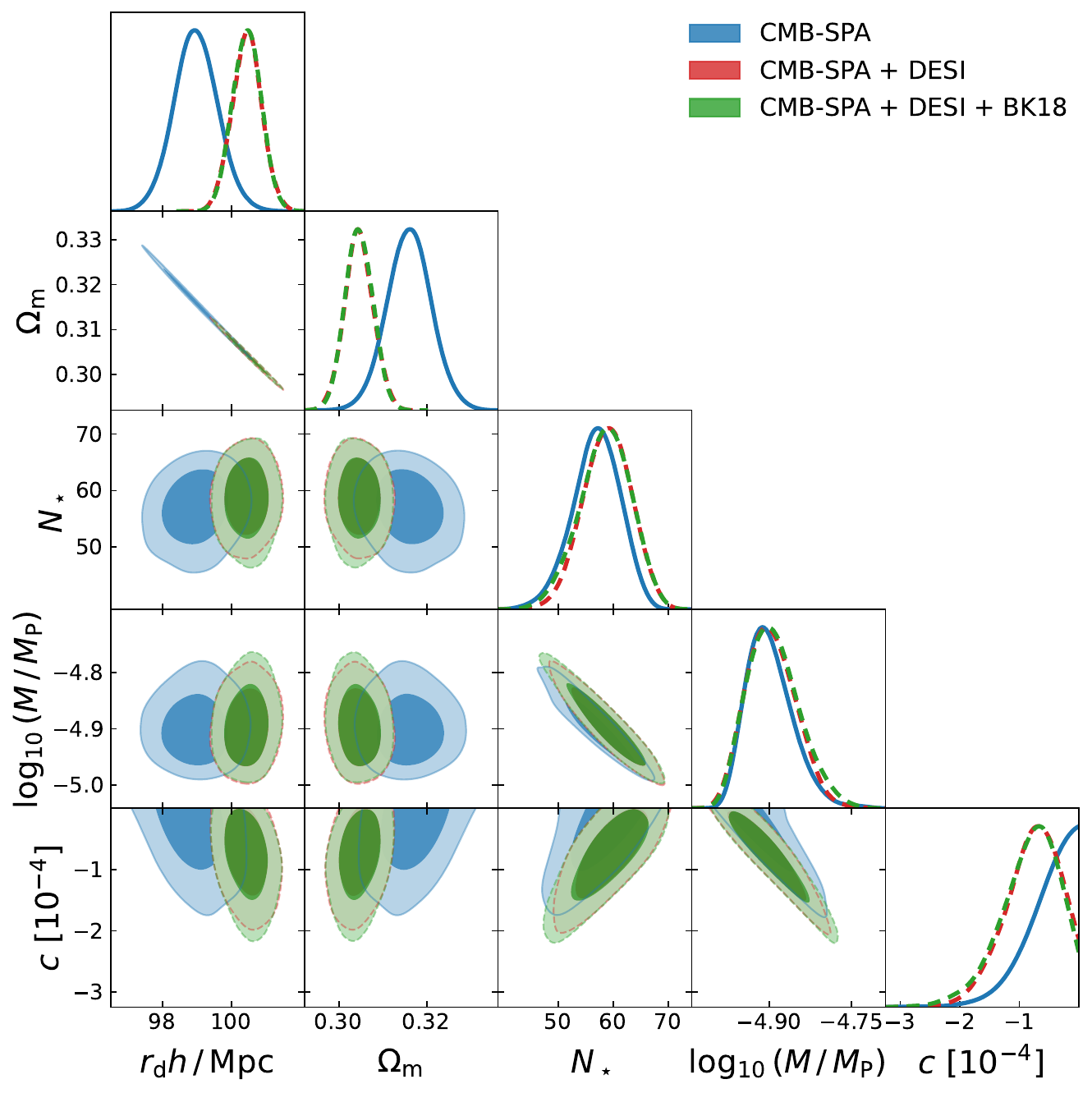}
  \caption{Same as Fig.~\ref{fig:rdplot} but for $R^3$ modified Starobinsky model with $c > 0$ (left) and $c<0$.}
  \label{fig:rdplotR3}
\end{figure}

\begin{table}[htbp!]
\centering
\renewcommand{\arraystretch}{1.25}
\resizebox{\textwidth}{!}{%
\begin{tabular}{|l|c|c|c|c|c|c|}
\hline
& \multicolumn{2}{c|}{\textbf{CMB-SPA}}
& \multicolumn{2}{c|}{\textbf{CMB-SPA+DESI}}
& \multicolumn{2}{c|}{\textbf{CMB-SPA+DESI+BK18}} \\[1.2ex]
\cline{2-7}
\textbf{$R^3$, $c >0$} &
\textbf{Mean $\pm$ 68\% CI} & \textbf{Best-fit} &
\textbf{Mean $\pm$ 68\% CI} & \textbf{Best-fit} &
\textbf{Mean $\pm$ 68\% CI} & \textbf{Best-fit} \\[0.8ex]
\hline
$M\ [10^{-5}]$ & $1.042^{+0.097}_{-0.086}$ & 0.998 & $1.025^{+0.066}_{-0.068}$ & 1.037 & $1.022^{+0.072}_{-0.066}$ & 1.078 \\
$c\ [10^{-5}]$ & $3.903^{+0.984}_{-3.903}$ & 5.012 & $2.031^{+0.418}_{-2.031}$ & 0.311 & $2.031^{+0.437}_{-2.031}$ & 0.019 \\
$N_*$ & $62.45^{+3.87}_{-4.57}$ & 63.78 & $65.14^{+3.71}_{-3.62}$ & 65.46 & $65.38^{+3.43}_{-4.10}$ & 63.43 \\
\hline
\hline
\cline{2-7}
\textbf{$R^3$, $c <0$} &
\textbf{Mean $\pm$ 68\% CI} & \textbf{Best-fit} &
\textbf{Mean $\pm$ 68\% CI} & \textbf{Best-fit} &
\textbf{Mean $\pm$ 68\% CI} & \textbf{Best-fit} \\[0.8ex]
\hline
$M\ [10^{-5}]$ & $1.270^{+0.083}_{-0.151}$ & 1.278 & $1.277^{+0.105}_{-0.156}$ & 1.278 & $1.290^{+0.103}_{-0.172}$ & 1.269 \\
$c\ [10^{-5}]$ & $-5.462^{+5.462}_{-1.138}$ & -5.972 & $-7.991^{+5.770}_{-3.300}$ & -7.501 & $-8.444^{+6.097}_{-3.321}$ & -7.835 \\
$N_*$ & $56.88^{+4.70}_{-3.94}$ & 56.49 & $58.84^{+4.46}_{-4.45}$ & 58.18 & $58.49^{+5.13}_{-4.27}$ & 58.73 \\
\hline
\end{tabular}}
\caption{The best-fit and mean $\pm$ 68\% CI for the model parameters for $R^3$ modified Starobinsky inflation for the CMB-SPA, CMB-SPA+DESI, and CMB-SPA+DESI+BK18 combinations, with the upper and lower halves displaying the $c > 0$ and $c < 0$ branches respectively.}
\label{tab:r3}
\end{table}
The mean $\pm$ 68\% and best-fit values of $N_*$ for the CMB-SPA, CMB-SPA+DESI, and CMB-SPA+DESI+BK18 combinations are $62.45^{+3.87}_{-4.57}$, $65.14^{+3.71}_{-3.62}$, and $65.38^{+3.43}_{-4.10}$, with best-fit values $63.78$, $65.46$, and $63.43$, respectively, for $c > 0$, whereas for $c < 0$ they are $56.88^{+4.70}_{-3.94}$, $58.84^{+4.46}_{-4.45}$, and $58.49^{+5.13}_{-4.27}$, with best-fit $56.49$, $58.18$, and $58.73$. The mean $\pm$ 68\% values of $N_*$ for $c < 0$ should be compared with the ones from Refs.~\cite{Gialamas:2025ofz,Kim:2025dyi,Addazi:2025qra,Park:2025upd}. We find a similar trend, but the $N_*$ and $c$ values are somewhat different from those of Refs.~\cite{Gialamas:2025ofz,Kim:2025dyi,Addazi:2025qra,Park:2025upd}, which may arise due to our approach where we directly solve the EoMs to find the posterior. We further remark that the result also depends on the choice of priors for the model parameters. In particular, adopting a broader prior on $N_*$ slightly weakens the resulting constraint. Further, in Fig.~\ref{fig:rdplotR3}, we also display the correlation of the model parameter with the $r_d h$ and $\Omega_m$. The model parameters $M$, $N_*$ and $c$ exhibit similar tension as observed in Fig.~\ref{fig:rdplot}.

Unlike the baseline Starobinsky model, the presence of $c$ prevents us from obtaining a simple closed analytic form as in Eq.~\eqref{eq:nstar}, and hence a direct constraint on $T_{\rm re}$ as a derived parameter. Instead, we take the allowed values of $M$, $N_*$, and $c$, assume a particle content, and estimate $T_{\rm re}$ by imposing the matching condition in Eq.~\eqref{eq:kefrel} (see Refs.~\cite{Gonuguntla:2026rkw,Modak:2025grj,Modak:2025bjv} for similar discussions).

\section{The Slow-Roll Parametrizations}
\label{sec:hsr}
In the previous section, we presented constraints on the Starobinsky model and its extension within canonical single-field inflation framework. It is also instructive to analyze how the latest data constrain the effective Hubble and potential slow-roll parametrizations of inflation as model independent constraints. First, we reconstruct the Hubble function $H(\varphi)$ and the corresponding potential using the exact Hamilton-Jacobi relations,
\begin{align}
V(\varphi)
&= 3M_{\rm P}^2 H^2(\varphi)
 - 2M_{\rm P}^4 \left(\frac{\mathrm{d} H}{\mathrm{d} \varphi}\right)^2,
\end{align}
with the background field evolution determined by
\begin{align}
\dot{\varphi}=-2M_{\rm P}^2\frac{\mathrm{d} H}{\mathrm{d} \varphi}.
\end{align}
Although these relations are exact, only the observable portion of the inflationary potential can be reconstructed. Following Refs.~\cite{Lesgourgues:2007aa,Lesgourgues:2007gp,Planck:2018jri} (see also Refs.~\cite{Liddle:1994dx,Kinney:2002qn}), we expand the Hubble function around the pivot field value $\varphi_*$ as
\begin{align}
H(\varphi)
=\sum_{n=0}^{N}\frac{1}{n!}
\left.\frac{\mathrm{d}^nH}{\mathrm{d} \varphi^n}\right|_{\varphi_*}
(\varphi-\varphi_*)^n.
\label{eq:recohubble}
\end{align}
In order to avoid parameter degeneracy, it is more convenient to parameterize the reconstruction in terms of the Hubble slow-roll hierarchy~\cite{Lesgourgues:2007aa,Lesgourgues:2007gp,Planck:2018jri}, rather than sampling the Taylor coefficients directly, as
\begin{equation}
\epsilon_H \equiv -\frac{\dot{H}}{H^2}, \quad
\eta_H \equiv \frac{\dot{\epsilon}_H}{H\epsilon_H}, \quad
\xi_H^2 \equiv \frac{\dot{\eta}_H}{H\eta_H}, \quad
\omega_H^3 \equiv \frac{\dot{\xi}_H^2}{H\xi_H^2},
\label{eq:hsr_def}
\end{equation}
where overdots denote derivatives with respect to cosmic time. These parameters characterize the evolution of the Hubble expansion rate and, consequently, the local shape of the inflationary trajectory. The zeroth-order hierarchy is denoted by $A_s^{\rm HSR}$, with $\epsilon_H>0$. We sample ${A_s^{\rm HSR}, \epsilon_H,\eta_H,\xi_H^2,\omega_H^3}$ assuming flat priors, together with the standard $\Lambda$CDM parameters $\omega_{\rm b}$, $\omega_{\rm cdm}$, $h$, and $\tau_{\rm reio}$. We restrict ourselves to the hierarchy up to $n=4$ (i.e., fourth order), make no assumption about the end of inflation and follow the default definition implemented in CLASS while keeping $k_* = 0.05{\rm Mpc}^{-1}$

\begin{figure}[htbp!]
  \centering
  \includegraphics[width=0.5\textwidth]{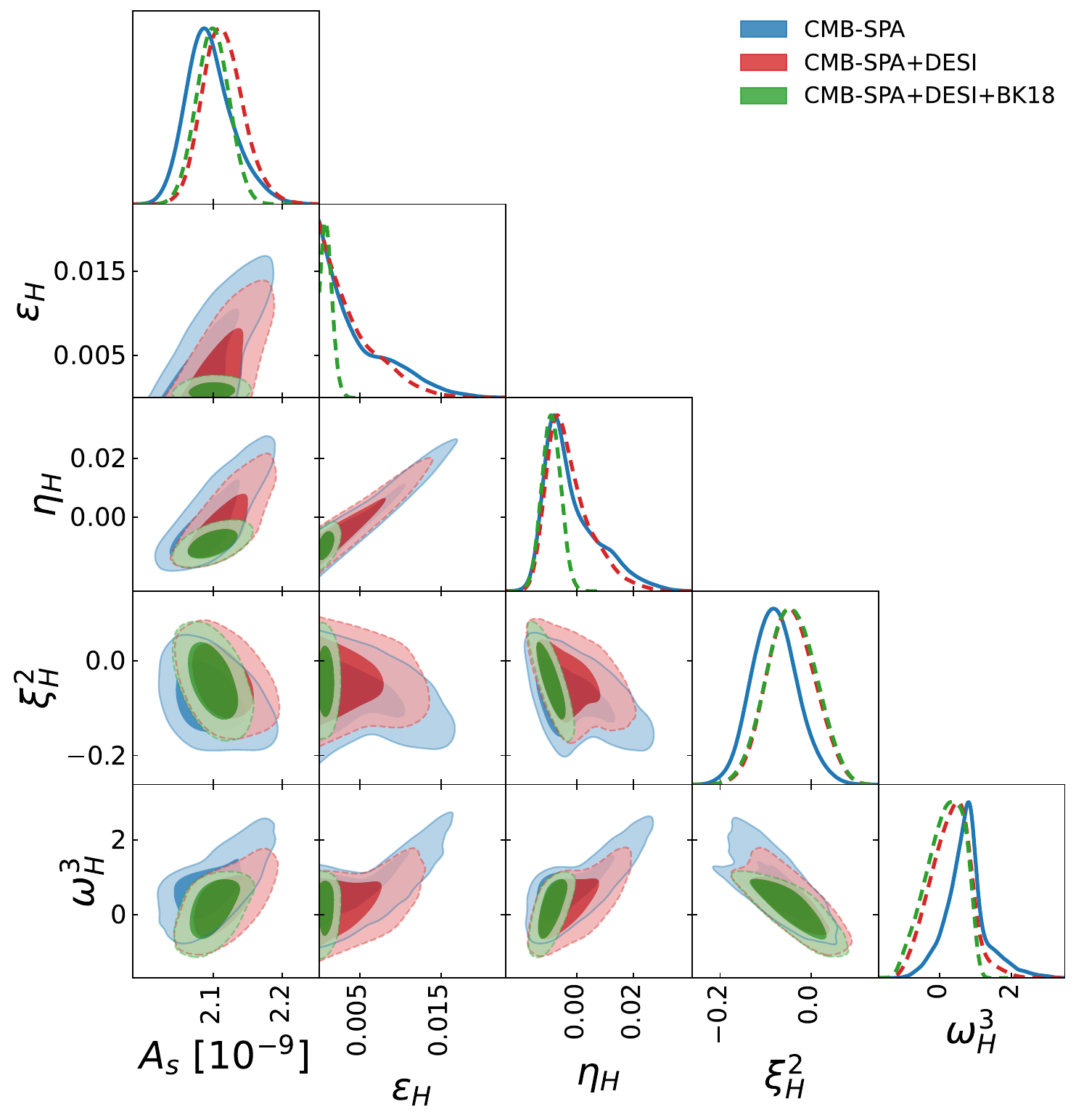}
  \caption{The constraints for HSR with 4th order hierarchy for the CMB-SPA, CMB-SPA+DESI and CMB-SPA+DESI+BK18 combinations. The darker and lighter shaded regions represent for 68\% and 95\% CI.}
  \label{fig:hsr_full}
\end{figure}

\begin{table}[htbp!]
\centering
\renewcommand{\arraystretch}{1.25}
\resizebox{\columnwidth}{!}{%
\begin{tabular}{|l|c|c|c|c|c|c|}
\hline
& \multicolumn{2}{c|}{\textbf{CMB-SPA}}
& \multicolumn{2}{c|}{\textbf{CMB-SPA+DESI}}
& \multicolumn{2}{c|}{\textbf{CMB-SPA+DESI+BK18}} \\[1.2ex]
\cline{2-7}
\textbf{Sampled} &
\textbf{Mean $\pm$ 68\% CI} & \textbf{Best-fit} &
\textbf{Mean $\pm$ 68\% CI} & \textbf{Best-fit} &
\textbf{Mean $\pm$ 68\% CI} & \textbf{Best-fit} \\[0.8ex]
\hline
$10^{9}A_s^{\rm HSR}$ & $2.096^{+0.027}_{-0.039}$ & 2.115 &
$2.115^{+0.026}_{-0.032}$ & 2.116 &
$2.099\pm0.024$ & 2.118 \\
$10^{3}\epsilon_H$ & $< 6.248 \; (13.51)$ & 4.924 &
$< 5.052 \; (11.03)$ & 4.100 &
$< 1.390 \; (2.204)$ & 0.655 \\
$10^{2}\eta_H$ & $-0.168^{+0.447}_{-1.187}$ & -0.210 &
$-0.255^{+0.435}_{-0.958}$ & -0.245 &
$-0.902^{+0.337}_{-0.335}$ & -0.783 \\
$\xi_H^2$ & $-0.0796^{+0.0462}_{-0.0535}$ & -0.0692 &
$-0.0437^{+0.0500}_{-0.0550}$ & -0.0443 &
$-0.0429^{+0.0540}_{-0.0533}$ & -0.0752 \\
$\omega_H^3$ & $0.737^{+0.398}_{-0.571}$ & 0.742 &
$0.304^{+0.581}_{-0.500}$ & 0.335 &
$0.149^{+0.663}_{-0.384}$ & 0.449 \\
\hline
\cline{2-7}
\textbf{Derived} &
\textbf{Mean $\pm$ 68\% CI} & \textbf{Best-fit} &
\textbf{Mean $\pm$ 68\% CI} & \textbf{Best-fit} &
\textbf{Mean $\pm$ 68\% CI} & \textbf{Best-fit} \\[0.8ex]
\hline
$10^9 A_s$ & $2.107^{+0.021}_{-0.022}$ & 2.128 &
$2.129\pm0.021$ & 2.130 &
$2.128^{+0.020}_{-0.021}$ & 2.141 \\
$n_s$ & $0.9661^{+0.0039}_{-0.0038}$ & 0.9666 &
$0.9727^{+0.0034}_{-0.0035}$ & 0.9728 &
$0.9726^{+0.0035}_{-0.0034}$ & 0.9721 \\
$r$ & $< 0.1138 \; (0.2249)$ & 0.0781 &
$< 0.0823 \; (0.1781)$ & 0.0650 &
$< 0.0219 \; (0.0356)$ & 0.0104 \\
$\alpha_s$ & $0.00551^{+0.00577}_{-0.00571}$ & 0.00355 &
$0.00397^{+0.00556}_{-0.00557}$ & 0.00359 &
$0.00531^{+0.00530}_{-0.00529}$ & 0.00776 \\
\hline
\end{tabular}}
\caption{The best-fit and mean for the HSR (4th order) parameters for different data combinations. For the strictly non-negative parameters ($10^3\epsilon_H$ and $r$), the statistics are presented as the $68\%$ one-tailed upper limit, with the approximate $95\%$ upper limit enclosed in parentheses.}
\label{tab:hsr}
\end{table}

The posterior for different HSR parameters are displayed in the left panel of Fig.~\ref{fig:hsr_full} for the CMB-SPA, CMB-SPA+DESI and CMB-SPA+DESI+BK18 datasets.\footnote{For validation, we have reproduced the Planck TT,TE,EE+lowE 2018 results, which we do not display here and instead refer the reader to Ref.~\cite{Modak:2022gol}.} The best-fit values and mean $\pm68\%$ CI for ${A_s^{\rm HSR}, \epsilon_H,\eta_H,\xi_H^2,\omega_H^3}$ are presented in Table~\ref{tab:hsr}. At this point, it is also useful to compare our results with those of Planck 2018~\cite{Planck:2018jri}. Planck TT,TE,EE+lowE+lensing+BK15 found $\epsilon_H < 0.0041$ at 95\% CL, whereas, adding our error bars in quadrature, we find $\epsilon_H \lesssim 0.0022$ at 95\% CL. This means the upper limit is improved by about $\sim 46\%$. For $\eta_H$, we find that the error bar has been reduced by $\sim 40\%$ compared to Planck TT,TE,EE+lowE+lensing+BK15, while in both cases $\eta_H$ exhibits a negative bias. For $\xi_H^2$, we find a negative bias (unlike Planck 2018), and for $\omega_H^3$, we find a positive bias, as in Planck 2018, but the uncertainty has shrunk by $\sim 27\%$ for $\xi_H^2$ and remains practically unchanged for $\omega_H^3$. We also provide constraints on the derived parameters $n_s$, $\alpha_s$, $r$ (with $r>0$), and $A_s$, obtained from the primordial spectra via the default implementation of CLASS, and find that $n_s$ and $A_s$ show trends similar to those of the baseline $\Lambda$CDM parameters across the datasets, as discussed in Ref.~\cite{SPT-3G:2025bzu}. We find that the derived constraints on the tensor-to-scalar ratio $r$ show a mild positive bias for CMB-SPA+DESI+BK18. This primarily reflects the single-field relation $r \approx 16\epsilon_H$; since $\epsilon_H$ exhibits a mild positive bias, so does $r$.

\begin{table}[htbp!]
\centering
\renewcommand{\arraystretch}{1.25}
\resizebox{\columnwidth}{!}{%
\begin{tabular}{|l|c|c|c|c|c|c|}
\hline
& \multicolumn{2}{c|}{\textbf{CMB-SPA}}
& \multicolumn{2}{c|}{\textbf{CMB-SPA+DESI}}
& \multicolumn{2}{c|}{\textbf{CMB-SPA+DESI+BK18}} \\[1.2ex]
\cline{2-7}
\textbf{Sampled} &
\textbf{Mean $\pm$ 68\% CI} & \textbf{Best-fit} &
\textbf{Mean $\pm$ 68\% CI} & \textbf{Best-fit} &
\textbf{Mean $\pm$ 68\% CI} & \textbf{Best-fit} \\[0.8ex]
\hline
$10^{9}A_s^{\rm PSR}$ & $2.098^{+0.035}_{-0.042}$ & 2.122 &
$2.109^{+0.033}_{-0.038}$ & 2.115 &
$2.086^{+0.027}_{-0.028}$ & 2.085 \\
$10^{3}\epsilon_V$ & $< 6.712 \; (11.50)$ & 4.176 &
$< 5.319 \; (10.59)$ & 2.288 &
$< 1.594 \; (2.293)$ & 0.701 \\
$10^{2}\eta_V$ & $0.938^{+0.963}_{-1.425}$ & 0.905 &
$0.439^{+0.789}_{-1.360}$ & 0.145 &
$-0.645^{+0.490}_{-0.553}$ & -0.810 \\
$10^{2}\xi_V^2$ & $-1.347^{+0.707}_{-0.722}$ & -1.580 &
$-0.564^{+0.612}_{-0.598}$ & -0.966 &
$-0.397^{+0.578}_{-0.560}$ & -0.373 \\
$10^{2}\varpi_V^3$ & $0.904^{+0.534}_{-0.537}$ & 1.062 &
$0.316^{+0.446}_{-0.454}$ & 0.517 &
$0.106^{+0.392}_{-0.381}$ & 0.109 \\
\hline
\cline{2-7}
\textbf{Derived} &
\textbf{Mean $\pm$ 68\% CI} & \textbf{Best-fit} &
\textbf{Mean $\pm$ 68\% CI} & \textbf{Best-fit} &
\textbf{Mean $\pm$ 68\% CI} & \textbf{Best-fit} \\[0.8ex]
\hline
$10^9 A_s$ & $2.108^{+0.021}_{-0.022}$ & 2.128 &
$2.130\pm0.021$ & 2.135 &
$2.128\pm0.020$ & 2.131 \\
$n_s$ & $0.9631\pm0.0047$ & 0.9636 &
$0.9719\pm0.0037$ & 0.9710 &
$0.9725^{+0.0036}_{-0.0035}$ & 0.9723 \\
$r$ & $< 0.1067 \; (0.1914)$ & 0.0666 &
$< 0.0880 \; (0.1720)$ & 0.0362 &
$< 0.0246 \; (0.0372)$ & 0.0110 \\
$\alpha_s$ & $0.00665^{+0.00559}_{-0.00561}$ & 0.00775 &
$0.00417^{+0.00546}_{-0.00548}$ & 0.00765 &
$0.00535^{+0.00522}_{-0.00519}$ & 0.00486 \\
\hline
\end{tabular}}
\caption{The best-fit and mean of the PSR (4th order) parameters with CMB-SPA, CMB-SPA+DESI and CMB-SPA+DESI+BK18 combinations. For the strictly non-negative parameters ($10^3\epsilon_V$ and $r$), the statistics are presented as the $68\%$ one-tailed upper limit, with the approximate $95\%$ upper limit enclosed in parentheses.}
\label{tab:psr}
\end{table}

\begin{figure}[htbp!]
  \centering
  \includegraphics[width=0.5\textwidth]{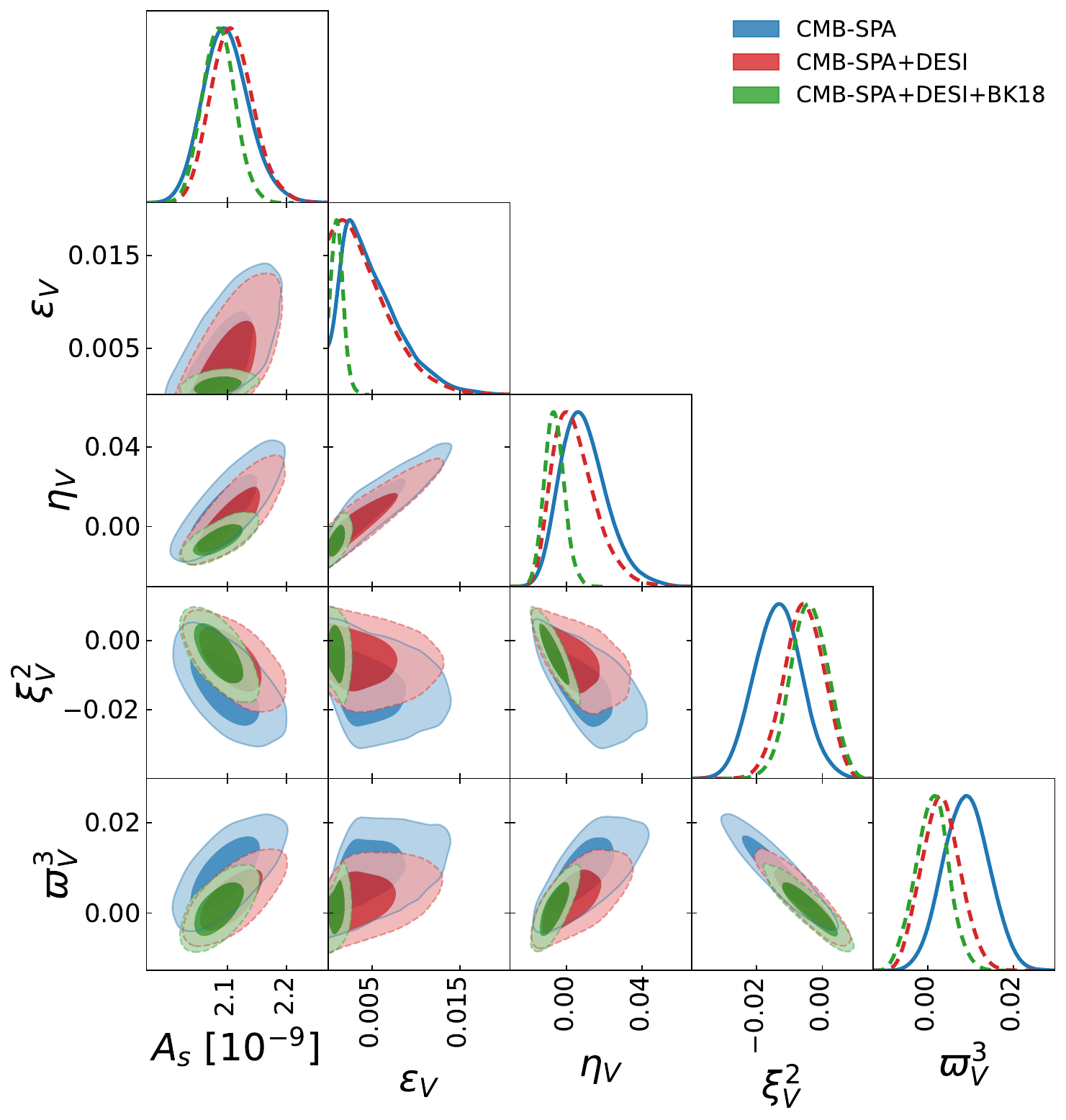}
  \caption{The constraints for PSR  with $n=4$ order hierarchy for the CMB-SPA, CMB-SPA+DESI and CMB-SPA+DESI+BK18 combinations.}
  \label{fig:psr_full}
\end{figure}

We now focus on the so-called potential slow-roll (PSR) parametrization. In this parametrization, one reconstructs the inflaton potential by Taylor-expanding $V(\varphi)$ around the pivot field value $\varphi=\varphi_*$, allowing the potential to be reconstructed within the observable window for canonical single-field inflation without making any assumption on the end of inflation. Again, instead of sampling the Taylor coefficients $\{V,V',V'',V''',\ldots\}$ directly, it is computationally more efficient to sample the corresponding PSR hierarchy, defined up to fourth order as~\cite{Planck:2018jri}
\begin{equation}
    \epsilon_V \equiv \frac{M_{\rm P}^2}{2}\left(\frac{V'}{V}\right)^2, \quad
    \eta_V \equiv M_{\rm P}^2\frac{V''}{V}, \quad
    \xi_V^2 \equiv M_{\rm P}^4\frac{V'V'''}{V^2}, \quad
    \varpi_V^3 \equiv M_{\rm P}^6\frac{(V')^2V''''}{V^3},
    \label{eq:psr_def}
\end{equation}
where primes denote derivatives with respect to $\varphi$ with $\epsilon_V>0$. We sample the parameter set $\{A_s^{\rm PSR}, \epsilon_V, \eta_V, \xi_V^2, \varpi_V^3\}$ with flat priors with $\omega_{\rm b}$, $\omega_{\rm cdm}$, $h$, and $\tau_{\rm reio}$. We stress that this choice of prior does not imply any slow-roll approximation in the calculation of the primordial spectra and no assumption.

The corresponding bounds of the PSR parameters with $n=4$ order are presented in Fig.~\ref{fig:psr_full} and the corresponding best-fit and mean values are summarized in Table~\ref{tab:psr}. We find the CMB-SPA+DESI+BK18 improves the constraint on $\epsilon_V$ by $\gtrsim 52\%$ compared to Planck TT, TE, EE+lowE+lensing+BK15. We find the upper limit $\epsilon_V \lesssim 0.0023$ compared to $< 0.0048$ at 95\% CL from Planck TT, TE, EE+lowE+lensing+BK15. The errors on higher order parameters $\eta_V$, $\xi^2_V$ and $\varpi^3_V$ are improved by $\sim 52\%$, $\sim 43\%$ and $\sim 35\%$. We find that $\eta_V$ and $\xi^2_V$ have negative but $\varpi^3_V$ has positive bias as in Planck TT, TE, EE+lowE+lensing+BK15. We also find a similar trend for the derived parameters $n_s$, $\alpha_s$, $r$ (with $r>0$), and $A_s$, here as well, i.e. for the data combination with DESI, the $n_s$ is pushed towards higher value.

Note here we have not shown the constraints on $\omega_{\rm b}$, $\omega_{\rm cdm}$, $h$, and $\tau_{\rm reio}$ along with the PSR and HSR as they show similar features as in above.

\section{Summary and Outlook}\label{sec:disc}
We analyze the sensitivity of combined CMB data from Planck, ACT, and SPT, together with BAO data from DESI, to constrain the baseline Starobinsky model without assuming the slow-roll approximation. For three data combinations, CMB-SPA, CMB-SPA+DESI, and CMB-SPA+DESI+BK18, we derive constraints on the Starobinsky model parameters $M$ and $N_*$. We find that the inferred values of $N_*$ depend significantly on the data combination. For CMB-SPA alone, the Starobinsky model remains the best-fit model, with $N_*=62.82$. Including DESI BAO data shifts the best-fit value to $N_*=71.06$, which decreases slightly to $N_*=69.94$ when BK18 data are further included. These values are consistently higher than those inferred from the single-field slow-roll relation $n_s\simeq 1-2/N_*$ using the marginalized mean of $n_s$ in the $\Lambda$CDM model. We further discuss the parameter degeneracy between $M$ and $N_*$ and the impact of the prior choices. Finally, we examine the correlations among $M$, $N_*$, and $r_dh$ and find effects similar to those reported in Ref.~\cite{SPT-3G:2025bzu}.

The parameter constraints of the baseline Starobinsky model can be recast as constraints on the reheating temperature under the slow-roll approximation. For $N_{\rm re}=0,2,4$, we find $T_{\rm re}\approx 1.64\times 10^{12}$ GeV, $2.22\times 10^{11}$ GeV, and $3.01\times 10^{10}$ GeV, respectively, for CMB-SPA. Although the best-fit value is higher, the reheating temperature predicted assuming Standard Model particle content with minimal gravitational coupling, $T_{\rm re}\sim10^8$--$10^9$ GeV, lies within the corresponding 68\% CI. For CMB-SPA+DESI, the best-fit values are $T_{\rm re}\approx 3.86\times 10^8$ GeV, $5.22\times 10^7$ GeV, and $7.07\times 10^6$ GeV for $N_{\rm re}=0, 2,$ and $4$, respectively. For CMB-SPA+DESI+BK18, the corresponding best-fits are $1.21\times 10^9$ GeV, $1.63\times 10^8$ GeV, and $2.21\times 10^7$ GeV. For CMB-SPA+DESI and CMB-SPA+DESI+BK18, we find lower best-fit reheating temperatures for the respective $N_{\rm re}$ values than those obtained from CMB-SPA alone. This behavior can be understood from Eqs.~\eqref{eq:nstar} and \eqref{eq:slow}, together with the corresponding constraints on $N_*$.

We also discuss the parameter constraints for the $R^3$-modified Starobinsky model. We find that, in the case of the $c > 0$ branch, the marginalized mean of $N_*$ is higher than that of the $c < 0$ branch across the data combinations. For instance, with CMB-SPA, the mean is $N_* = 62.45$ for $c > 0$, compared to $N_* = 56.88$ for $c < 0$ and with CMB-SPA+DESI, the mean is $N_* = 65.14$ for $c > 0$, compared to $N_* = 58.84$ for $c < 0$. This pattern persists when BK18 data is added, yielding $N_* = 65.38$ for $c > 0$ and $N_* = 58.49$ for $c < 0$. In both cases, the $N_*$ means are notably different from those estimated using the single-field slow-roll approximation through $n_s$.

Finally, we provide the parameter constraints for the HSR and PSR parametrizations up to fourth order in the slow-roll hierarchy. The resulting constraints are consistent with the baseline $\Lambda$CDM model, as reflected by the derived values of $n_s$ and $A_s$ in Tables~\ref{tab:hsr} and \ref{tab:psr}. These model-independent constraints can be readily applied to any inflationary potential satisfying the slow-roll assumption.

While the high $n_s$ may reflect unaccounted-for uncertainties, more precise measurements are needed to resolve the tension. Future CMB data from the Simons Observatory, expected to reach $\sigma(n_s)\sim0.002$~\cite{SimonsObservatory:2025wwn}, together with BAO measurements from DES~\cite{DES:2024pwq}, may help clarify the discrepancy.

\subsection*{Acknowledgments}

We thank Tilman Plehn for discussions.

\renewcommand{\emph}{}

\bibliography{references}


\end{document}